\documentclass[onecolumn,authoryear]{els-mrw} 

\usepackage{amsmath,amssymb,amsfonts,amsthm,makeidx,graphicx}
\usepackage{txfonts}
\usepackage{helvet}
\usepackage{aas_macros}
\usepackage{listings}
\usepackage{hyperref}
\usepackage{xcolor}
\hypersetup{colorlinks = True}
\usepackage{enumitem}
\usepackage{tcolorbox}

\usepackage{totcount}
\newtotcounter{citnum} 
\def\oldbibitem{} \let\oldbibitem=\bibitem
\def\bibitem{\stepcounter{citnum}\oldbibitem}

\begin{document}
\chapter{Main-sequence systems: orbital stability in stellar binaries}\label{chap1}

\author[1]{Billy Quarles}%
\author[2,3]{Hareesh Gautham Bhaskar}
\author[2]{Gongjie Li}%

\address[1]{\orgname{Texas A\&M University-Commerce}, \orgdiv{Department of Physics \& Astronomy}, \orgaddress{P.O. Box 3011 Commerce, TX 75429-3011, USA}}
\address[2]{\orgname{Georgia Institute of Technology}, \orgdiv{School of Physics}, \orgaddress{Howey Physics Bldg, 837 State St NW, Atlanta, GA 30332}}
\address[3]{\orgname{Technion - Israel Institute of Technology}, \orgdiv{Department of Physics}, \orgaddress{Haifa 3200003 Israel}}

\articletag{Chapter Article tagline: update of previous edition,, reprint..}

\maketitle





\begin{abstract}[Abstract]
The majority of star formation results in binaries or higher multiple systems, and planets in such systems are constrained to a limited range of orbital parameters in order to remain stable against perturbations from stellar companions. Many planets have been discovered in such multiple systems (such as stellar binaries), and understanding their stability is important in exoplanet searches and characterization. In this chapter, we focus on the orbital stability of planets in stellar binaries. We review key results based on semi-analytical secular (long term) methods, as well as results based on N-body simulations and more recent Machine Learning methods. We discuss planets orbiting one of the stellar binary components (S-type) and those orbiting both stars (P-type) separately.
\end{abstract}

\begin{tcolorbox}
\textbf{Key Points} \\
\term{Laplace-Lagrange} A Mathematical framework used in celestial mechanics to describe the long-term evolution of the orbits of planets and other celestial bodies, considering the gravitational perturbations they exert on each other.\\ 
\term{Disturbing function} A potential function used in celestial mechanics to quantify the perturbative gravitational effects of one celestial body on the orbit of another. \\
\term{Upper Critical Orbit} The highest stable orbit around a star beyond which the planet will experience destabilizing forces, leading to its eventual escape.\\
\term{Lower Critical Orbit} The lowest stable orbit around a star, below which a planet's orbit will become unstable.\\
\term{Barycentric coordinates} Coordinate system that represents the positions of celestial bodies relative to the center of mass (barycenter) of a system of bodies.\\ 
\term{CRTBP} Circular restricted three-body problem, a simplified model that studies the motion of a small body under the gravitational influence of two larger bodies (primary and secondary) which are in circular orbits.\\
\term{Lagrange equilibrium points} Positsions in a two-body system where a small third body can remain in a stable or unstable equilibrium due to the gravitational forces of the two primary bodies.\\
\term{Symplectic integration} A numerical method used to solve Hamiltonian systems in physics and celestial mechanics, which preserves the symplectic structure (e.g., conservation properties) of the system over long time periods.\\
\term{Hamilton's equations} A set of first-order differential equations that describe the time evolution of a physical system in classical mechanics using coordinates and momenta, derived from the Hamiltonian function, which represents the total energy of the system.\\
\term{Recall accuracy} A measure in classification tasks that quantifies the proportion of true positive instances correctly identified by the model out of all actual positive instances.\\
\term{Mean motion resonances} MMRs are integer ratios between the orbital periods of two planets. \\
\term{Hill radius} Characteristic radius around planets which their own gravity dominates relative to the gravity of the central star. \\
\term{Secular resonances} Resonances that correspond to precession of the orbital planes of the planets.\\
\term{S-type} Planet orbits either of the stellar binary components.\\
\term{P-type} Planet orbits both of the stellar binary components. \\
\term{Forced eccentricity} Eccentricities driven by planet-planet interactions. \\
\term{Free eccentricity} Eccentricities derived from the orbital boundary conditions.
\end{tcolorbox}


\section{Introduction}\label{chap1:sec3}
Observational surveys uncovered that nearly half of all Solar-type star systems (e.g., singles, binaries, triples, etc.) are actually stellar binaries \cite{Moe2017}.  Prior to such discoveries, considering the potential orbital stability of planets in binary systems appeared to be a hypothetical or mathematical curiosity.  But no longer, as ${\gtrsim}200$ confirmed exoplanets exist within binary star systems\footnote{\url{https://adg.univie.ac.at/schwarz/multiple.html}}, wherein ${\sim}10\%$ of these systems the planets completely orbit both stars. General studies for orbital stability of these planetary systems must consider the next step where the gravitational potential is no longer dominated by a central mass. 
\cite{Dvorak1982} coined the nomenclature for considering the different orbital types that may exist within a stellar binary. There are broadly two different types, where a planet may orbit:

\begin{itemize}
    \item either of the two stars like a \textbf{s}atellite (S-type), or
    \item both of the two stars like a \textbf{p}lanet (P-type). 
\end{itemize}

There is a third type, but it is much more specific because it is limited to low mass secondary companions (${\lesssim}5\%$ of the total binary mass).  This type describes planets in the vicinity of the equilibrium \textbf{L}agrange points $L_4$ or $L_5$ (L- or T-type).  Some investigations \citep[e.g.,][]{Lohinger1993,Schwarz2014} have explored T-Type planets in detail, where this is beyond our scope.

Planetary discoveries have caught up with theoretical developments, where one of the first exoplanet candidates $\gamma$ Cephei Ab \citep{Campbell1988} exists in a S-type configuration.  The Kepler Space Telescope observed ${\sim}12$ P-type systems (i.e., circumbinary planets or Tatooines), where the first confirmed system, Kepler-16 b, was identified soon after major science operations began \citep{Doyle2011}.

The orbital stability of a planet within a binary system depends on the gravitational force applied by the stellar companion, which depends on the masses and the relative distance between the stellar components.  The gravitational force is mediated by the universal gravitational constant $\mathcal{G}$, where its value can be defined by Newton's version of Kepler's third law and appropriate units.  For example, the masses of all three bodies (2 stars + 1 planet) can be measured in solar units ($M_\odot$) and, distances between each body in astronomical units (au), and the time to complete an orbit as $2\pi$, which together allows us to define $\mathcal{G}\equiv 1$.  Note that our choice for units is arbitrary as long as Newton's version of Kepler's third law is satisfied, where we could have chosen $\mathcal{G}\equiv 4\pi^2$ so that the distance is in au, time is in years, and mass is in solar masses.

Exoplanet orbital stability in stellar binaries more precisely defined means that the planet remains bound to the system for the entire stellar lifetime, where 10 Gyr is often used as a practical upper limit.  Additionally, the planet must avoid close approaches with either of the stellar components because such instances can either provide the gravitational assist required to liberate the planet or destroy the planet through an immense tidal force.  The planet's orbital elements can change over time as long as the system's angular orbital momentum remains conserved.

Long-term simulations for billions of years were untenable until the development of numerical integration schemes were developed specifically for planets in binary systems \citep{Chambers2002}.  Around the same time, theoretical studies of orbital stability advanced to use chaos indicators based on the Lyapunov characteristic number \citep[e.g.,][]{PilatLohinger2003} or the mean exponential growth of nearby orbits MEGNO \citep{Cincotta1999}.  In particular, the MEGNO was used to characterize the parameter space surrounding known exoplanets in binaries as either chaotic or regular \citep[e.g.,][]{Satyal2013}.  While regular orbits correlate with stable orbits in a straightforward manner, chaotic orbits can be either stable or unstable.  Therefore, methods using chaos indicators should be used with caution so that erroneous inferences are avoided.  In \S \ref{sec:Stype} and \ref{sec:Ptype}, we examine the historical and modern approaches to probe the orbital stability of an exoplanet within stellar binary systems.


\section{S-type stability}\label{sec:Stype}
\begin{figure}
    \centering
    \includegraphics[width=0.8\linewidth]{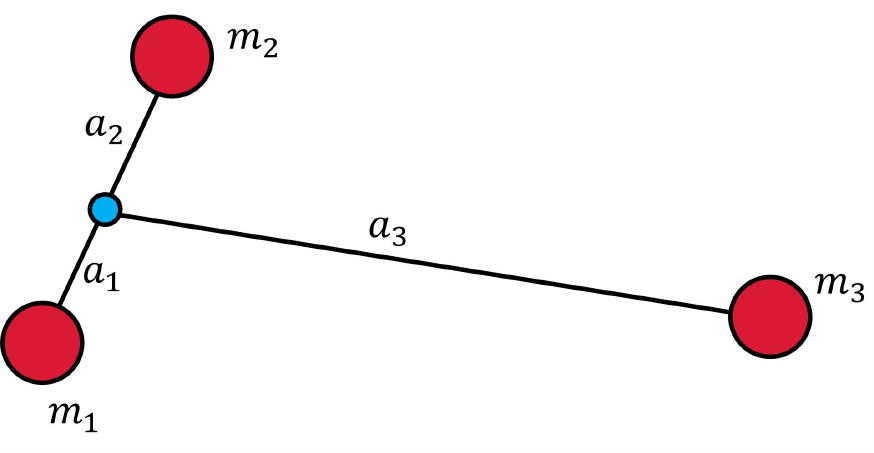}
    \caption{Orbital architecture of a planet on an S-Type orbit within a stellar binary, where the most massive component is $m_1$.  The planetary and secondary star mass is represented by $m_2$ and $m_3$, respectively.  The blue dot represents the center-of-mass for the inner pair, where planetary semimajor axis $a_{p} = a_1 + a_2$ and the stellar semimajor axis $a_{\rm bin} = a_3$.  }
    \label{fig:arch}
\end{figure}

For an S-Type orbit, let us consider the planet hosting star (primary) having a mass $m_1$, the planetary body has a mass $m_2$, and the other stellar (secondary) component has a mass $m_3$.  We necessarily define that $m_2\ll m_1$ and $m_1 \sim m_3$.  The planet and its host star orbit their common center-of-mass with semimajor axes $a_2$ and $a_1$, respectively.  From the conditions on the masses, the planetary semimajor axis can be written as $a_p = a_1 + a_2$, where $a_2\gg a_1$.  The secondary star has its own semimajor axis $a_3 = a_{\rm bin}$ relative to the inner pair's center-of-mass.  Figure \ref{fig:arch} illustrates this orbital architecture with each mass as a red dot, and the center-of-mass as a blue dot.

The three-body problem has no general solution \citep{Poincare1892}, but we can make certain approximations due to the nature of our problem.  The planetary mass is very small compared to either of the stellar masses so that we can apply a perturbation theory, which has been developed by \cite{Heppenheimer1978}, \cite{Marchal1990}, and \cite{AndradeInes2017} with varying complexity.  Such analytical approaches are still useful despite the ubiquity of numerical simulations because they provide a more complete view and insights towards how the three-body system may change under slightly different initial conditions.  

\subsection{Secular evolution through a disturbing function}\label{sec:Hepp}

In the hierarchical case ($\alpha = (a_p/a_{\rm bin}) \ll 1$), we can use the model by \cite{Heppenheimer1978}, which assumes there are no significant effects due to mean motion resonances.  In this case the planet's orbit will be altered by secular processes, which means that the eccentricity and/or orientation of the orbit can change, but not its semimajor axis (i.e., $\dot{a}_p \approx 0$).  The secular interaction with the perturber $m_3$ is modeled through a Laplace-Lagrange disturbing function, which can be expanded in terms of the semimajor axis ratio $\alpha$ using Legendre polynomials $(P_i)$, as $\alpha$ is a small parameter. In this model, we limit the expansion in Legendre polynomials to $P_3$ (i.e., quadrupole problem), which assumes that the planet's eccentricity $e_p\ (= e_2)$ is restricted to low values.  \cite{Heppenheimer1978} obtained the following secular disturbing function in orbital elements:

\begin{align} \label{eqn:S_disturb}
\mathcal{R}_{\rm Hep} &= \frac{\mathcal{G}m_3}{\left(1-e_{\rm bin}^2\right)^{3/2}} \left(\frac{a_p}{a_{\rm bin}}\right)^3 \left[ \frac{1}{4} + \frac{3}{8}e_p^2 - \frac{15}{16}\frac{a_p}{a_{\rm bin}}\frac{e_p e_{\rm bin}}{\left(1-e_{\rm bin}^2\right)}\cos(\Delta\varpi)\right],
\end{align}

where the subscript $p$ and ${\rm bin}$ refer to the planetary and binary orbital elements respectively.  The relative longitude $\Delta \varpi\ (=\varpi_p - \varpi_{\rm bin})$ represents the relative orientation between the planetary and binary orbits within a reference plane.  For simplicity, we assume that the stellar binary orbit is aligned along the $x$-axis so that $\varpi_{\rm bin}=0^\circ$.

In order to make the equations of motion linear and easier to solve, we introduce the planet's eccentricity vector components as

\begin{align} \label{eqn:ecc_vec}
\begin{aligned}
k &= e_p \cos\Delta\varpi, \\
h &= e_p \sin\Delta\varpi, 
\end{aligned}
\end{align}

and using the planet's mean motion $n_p = \sqrt{\mathcal{G}(m_1+m_2)/a_p^3}$, and the semimajor axis ratio $\alpha = a_p/a_{\rm bin}$, we can rewrite Eq. \ref{eqn:S_disturb} as

\begin{align}
\mathcal{R}_{\rm Hep} &= \frac{n_p^2a_p^2 \alpha^3}{\left(1-e_{\rm bin}^2\right)^{3/2}}\left(\frac{m_3}{m_1+m_2}\right) \left[ \frac{1}{4} + \frac{3}{8}\left(h^2 + k^2\right) - \frac{15}{16}\frac{ \alpha k e_{\rm bin} }{\left(1-e_{\rm bin}^2\right)}\right],
\end{align}

so that we can apply Hamilton's equations:

\begin{align}\label{eqn:hamilton}
\begin{aligned}
\dot{k} &= -\frac{1}{n_p a_p^2}\frac{\partial R_{\rm Hep}}{\partial h}, \\
\dot{h} &= \frac{1}{n_p a_p^2}\frac{\partial R_{\rm Hep}}{\partial k}.
\end{aligned}
\end{align}

As a result, we find

\begin{align}
\begin{aligned}
\dot{k} & = -\frac{3}{4}\frac{n_p \alpha^3}{\left(1-e_{\rm bin}^2\right)^{3/2}}\left(\frac{m_3}{m_1+m_2}\right) h = -g_H h, \\
\dot{h} &= \frac{3}{4}\frac{n_p \alpha^3}{\left(1-e_{\rm bin}^2\right)^{3/2}}\left(\frac{m_3}{m_1+m_2}\right) \left[  k - \frac{5}{4}\frac{ \alpha e_{\rm bin} }{\left(1-e_{\rm bin}^2\right)}\right] = g_H\left[  k - \epsilon_H \right],
\end{aligned}
\end{align}

where $g_H$ represents a constant frequency and $\epsilon_H$ is an offset to the $k$ component of the eccentricity vector.

Using the method to solve coupled first-order equations produces

\begin{align} \label{eqn:hk_SHO}
\begin{aligned}
\ddot{h} + g_H^2 h &= 0, \\
\ddot{k} + g_H^2 \left(k - \epsilon_H\right) &= 0.
\end{aligned}
\end{align}

Equation \ref{eqn:hk_SHO} has the form of a harmonic oscillator, which has the following analytical solutions: 

\begin{align} \label{eqn:forced_free}
\begin{aligned}
k(t) &= e_{\rm free}\cos\left(g_Ht + \phi\right) + \epsilon_H, \\
h(t) &= e_{\rm free}\sin\left(g_Ht + \phi\right),
\end{aligned}
\end{align}
where $e_{\rm free}$ and $\phi$ is determined by the initial conditions $k(0)$ and $h(0)$.  This method has introduced two additional parameters

\begin{align} \label{eqn:forced_free_params}
\begin{aligned}
g_H & = \frac{3}{4}\mu \alpha^3\frac{n_p }{\left(1-e_{\rm bin}^2\right)^{3/2}}, \\
\epsilon_H &= \frac{5}{4}\alpha\frac{  e_{\rm bin} }{\left(1-e_{\rm bin}^2\right)},
\end{aligned}
\end{align}

where $\mu$ is the mass ratio $m_3/(m_1+m_2) \approx m_3/m_1$, $g_H$ describes the precession frequency for a {\it free} eccentricity vector and $\epsilon_H$ is a stationary solution for the {\it forced} eccentricity vector induced from the stellar binary.  These two vectors combine to give the planetary eccentricity vector (see Eq. \ref{eqn:ecc_vec}), where $\Delta \varpi$ depends on time.  The values of $e_{\rm free}$ and $\phi$ can be determined from the magnitude difference and orientation of the planetary eccentricity vector relative to the forced eccentricity vector.  Mathematically, this is given by

\begin{align}
\begin{aligned}
e_{\rm free} &= \sqrt{\left[k(0) - \epsilon_H\right]^2 + \left[h(0)\right]^2}, \\
\tan \phi &= \frac{h(0)}{k(0) - \epsilon_H},
\end{aligned}
\end{align}
where the \texttt{arctan2} function should be used to find the value of $\phi$ in the proper quadrant.

\begin{figure}
    \centering
    \includegraphics[width=0.75\linewidth]{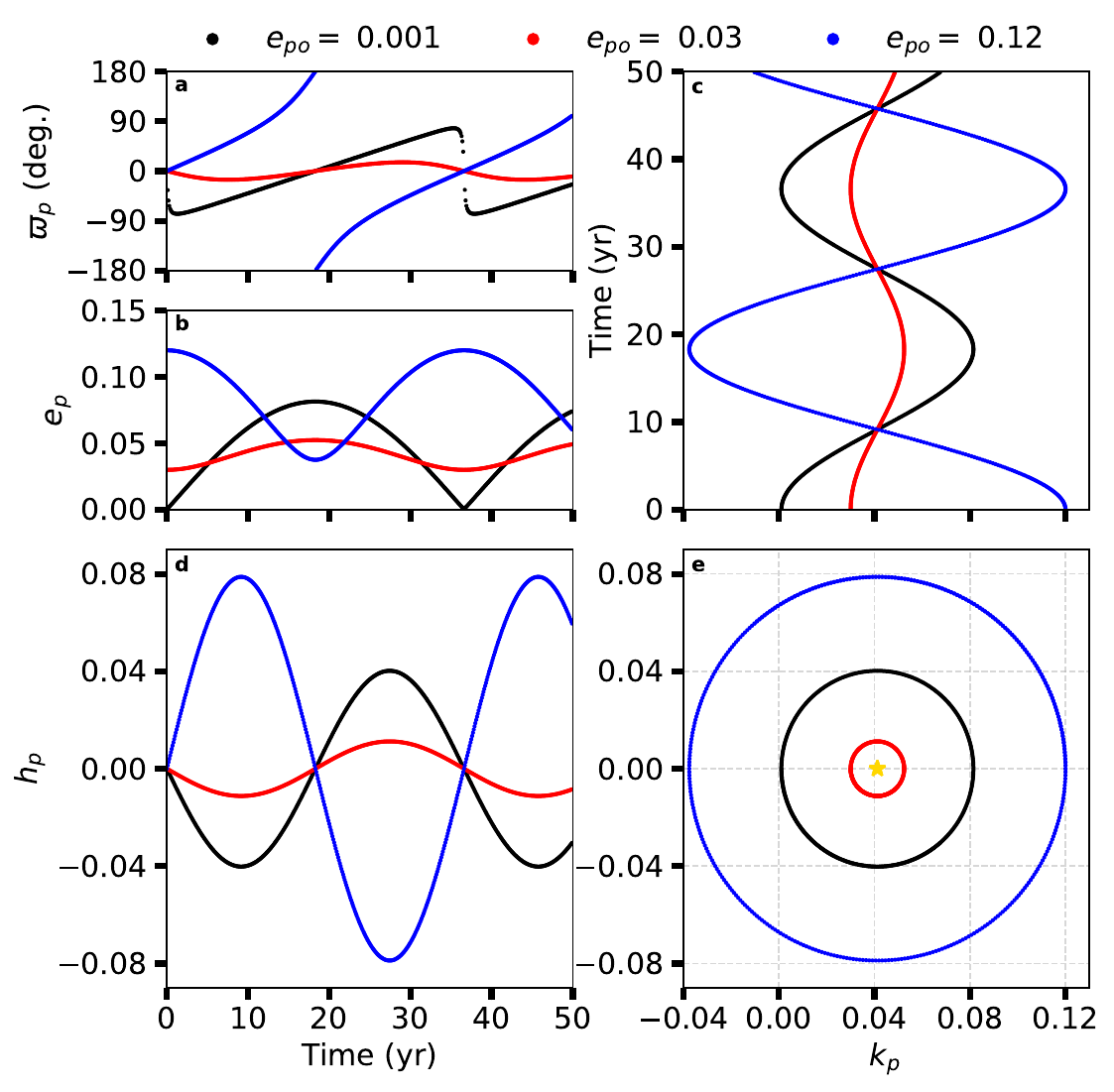}
    \caption{Secular time evolution for an S-Type planet using the \cite{Heppenheimer1978} model, where $m_1=m_3 = 1\ M_\odot$, $\alpha = 0.1$, $e_{\rm bin}=0.3$ for three different initial eccentricities $e_{po}$ initially aligned with the binary orbit $(\Delta\varpi = \varpi_{\rm bin} = 0^\circ)$.  The gold star in panel e represents the forced eccentricity $\epsilon_H$, which does not depend on $e_{po}$. Figure adapted from \citep{AndradeInes2017}.}
    \label{fig:Hep_sec}
\end{figure}

Figure \ref{fig:Hep_sec} illustrates the secular time evolution for three different initial planetary eccentricities $e_{po}$.  Note that $\epsilon_H$ and $g_H$ do not depend on the initial planetary eccentricity where the differing amplitudes in each panel Fig. \ref{fig:Hep_sec} depend on $e_{\rm free}$. Figs. \ref{fig:Hep_sec}a-\ref{fig:Hep_sec}d show that one complete cycle completes in 36.5 yr, which corresponds to $2\pi/g_H$.  The \citeauthor{Heppenheimer1978} model in the $(k,h)$ plane (Fig. \ref{fig:Hep_sec}e) shows the free eccentricity vector completing circular orbits with the end of the forced eccentricity vector as its origin.  Note that the red circle $(e_{po}=0.03)$ has the smallest radius and the black circle has a larger radius.

The \citeauthor{Heppenheimer1978} model has limitations, where it fails once its main assumptions are broken (e.g., $e_p \gtrsim 0.1$ or $\alpha \gtrsim 0.2$).  The Laplace-Lagrange formulation of the disturbing function is accurate to $\mathcal{O}(e_2^2)$, where \cite{Marchal1990} obtained a second-order solution in $\mu$ and $\alpha$ that is accurate to $\mathcal{O}(e_2^3)$ with the appropriate corrections to $g_H$ and $\epsilon_H$ \citep[see][]{Georgakarakos2005,AndradeInes2017}.  Although the disturbing function from \citeauthor{Marchal1990}'s model is move complicated, it can be simplified through the restricted problem $(m_p/m_1 \rightarrow 0)$ and ignoring the terms of $\mathcal{O}(e_p^3)$ to get

\begin{align} \label{eqn:S_Marchal}
\mathcal{R}_M = n_p a_p^2 g_M \left[\frac{1}{2}e_p^2 - \epsilon_M e_p \cos\Delta\varpi \right],
\end{align}
where

\begin{align}
g_M &= g_H(1+\delta_M), \\
\epsilon_M &= \epsilon_H(1 + \delta_M)^{-1}, \\
\delta_M &=  \frac{25}{8}\frac{\mu \alpha^{3/2}}{\sqrt{1+\mu}} \frac{3+2e_{\rm bin}^2}{(1-e_{\rm bin}^2)^{3/2}}.
\end{align}

Eventually, the \citeauthor{Marchal1990} approximation also breaks down, especially as $\alpha$ increases and other shorter period effects (e.g., mean motion resonances) are no longer negligible.  An alternative approach is to introduce {\it empirical} corrections to $g_H$ and $\epsilon_H$ from a reference numerical solution and applying a fitting procedure in terms of $\alpha$, $\mu$, and $e_{\rm bin}$.  \cite{Giuppone2011} demonstrated this method for the $\gamma$ Cephei system, where \cite{AndradeInes2017} generalized it to apply to eccentric binaries $(e_{\rm bin}\leq 0.6)$ and large mass ratios $(\mu \leq 10)$.  The resulting approximation was limited to $N_e = 15$ terms for the forced eccentricity correction and $N_g=18$ for the secular frequency.  Each approximation can be represented as a finite sum by

\begin{align}
\begin{aligned}
\delta_g &= \sum_{i=1}^{N_g} A_i^g \alpha^{p_i^g} \mu^{q_i^g} e_{\rm bin}^{l_i^g}, \\
\delta_e &= \sum_{i=1}^{N_e} A_i^e \alpha^{p_i^e} \mu^{q_i^e} e_{\rm bin}^{l_i^e},
\end{aligned}
\end{align}

where

\begin{align}
\begin{aligned}
g_{AI} &= g_H(1+\delta_g), \\
\epsilon_{AI} &= \epsilon_H(1+\delta_e).
\end{aligned}
\end{align}

The coefficients $A_i$ and corresponding exponents $(p_i,\ q_i,\ l_i)$ are tabulated in Appendix A of \cite{AndradeInes2017}. Note that when the mutual orbital inclination between the planet and the binary orbits is large $\gtrsim 40^\circ$, Von Zeipel-Lidov-Kozai oscillations can lead to large amplitude variations in planetary eccentricity and inclination (see discussions in section ``stability in hierarchical systems'' in chapter ``Main-sequence systems: orbital stability around single star hosts'').

\subsection{Applications of forced and free eccentricity} \label{sec:Stype_ecc}
The forced and free eccentricity provide a good indicator for the stability of an exoplanet on an S-Type orbit.  The forced eccentricity $\epsilon_H$ depends on the shape of the stellar binary's orbit $(a_{\rm bin}$ and $e_{\rm bin})$ and the planetary semimajor axis $a_p$, since $\alpha = a_p/a_{\rm bin}$.  This implies that the forced eccentricity sets a reference point for which the planetary eccentricity oscillates around (see Fig. \ref{fig:Hep_sec}).  To ensure the stability of the exoplanet, its free eccentricity vector must be such that the eccentricity variation (i.e., amplitude relative to the forced eccentricity) is minimized.

The perturbation theory approach (in Sect. \ref{sec:Hepp}) to determine the forced eccentricity is an approximation, whereas we may need a more accurate estimate.  In this case, we turn to numerical simulations that can efficiently forward-model a system given some initial parameters.  \texttt{Rebound} has become a standard option due to its versatility and user-friendly interface that can be easily implemented on browser-based \texttt{Jupyter} notebooks.

\cite{Quarles2018a} applied a combination of numerical and secular methods for understanding planetary stability in the $\alpha$ Centauri system $(m_A = 1.133\ M_\oplus$, $m_B = 0.972\ M_\oplus$ and $e_{\rm bin}=0.524)$.  Numerical n-body simulations provide a more complete solution, and in comparison to secular methods, we can see where the two begin to deviate.  Suppose there is a exoplanet orbiting $\alpha$ Cen B with a semimajor axis $a_p = 1.5$ au.  The stability of the exoplanet depends on its initial eccentricity $e_{po}$ and relative orbital alignment $\Delta \varpi$ to the binary orbit that define its initial eccentricity vector.  \cite{Quarles2018a} estimated, via n-body simulations, the forced eccentricity $e_F \approx 0.047$ using $\mu = 1.165$, where the \citeauthor{Marchal1990} secular approximation produces $\epsilon_M = 0.046$.  These two methods match with an accuracy more than $99.5\%$, where the secular approximation is far less expensive computationally.

\begin{figure}
    \centering
    \includegraphics[width=0.75\linewidth]{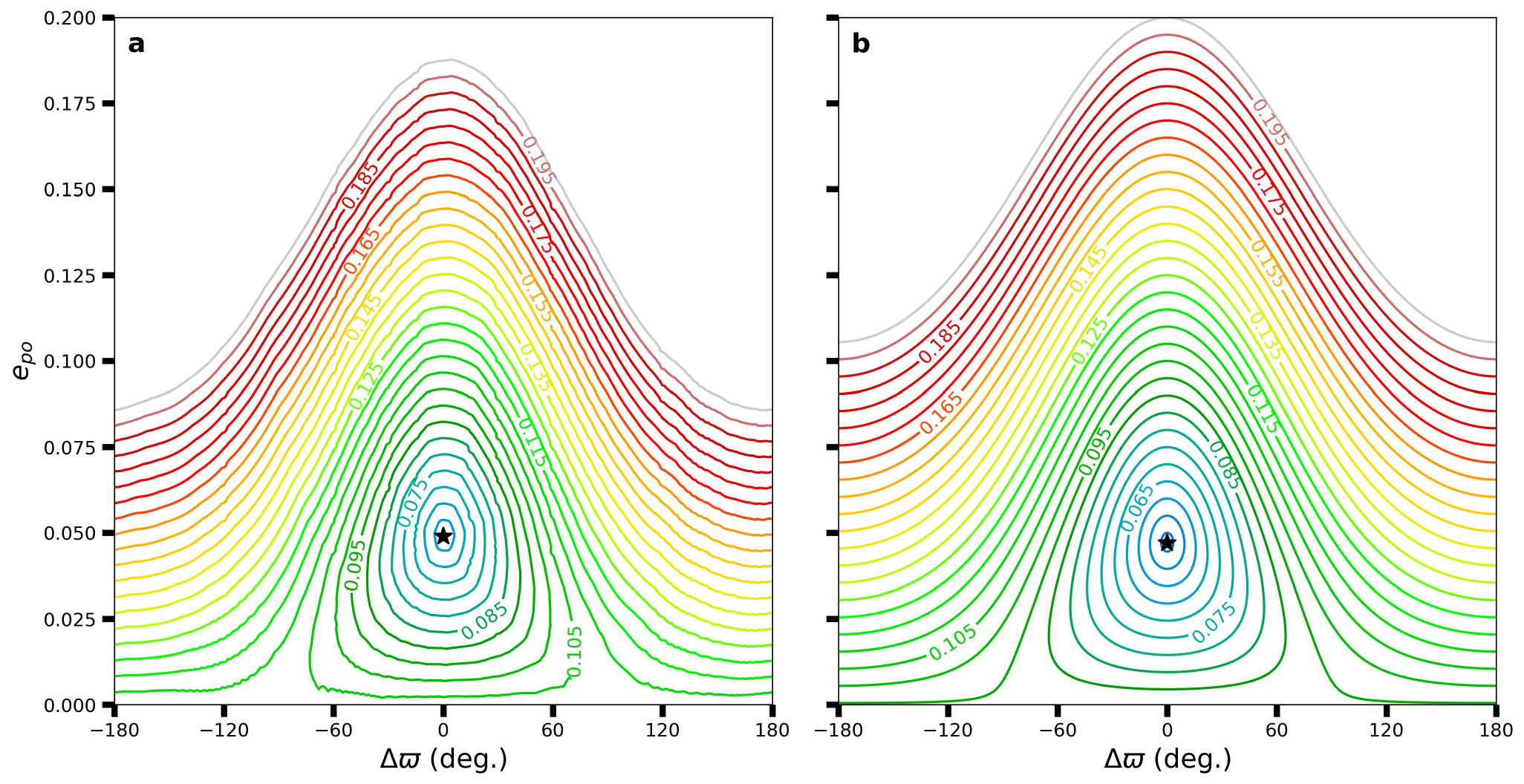}
    \caption{Maximum eccentricity (color-coded) for an {\bf S-Type} exoplanet orbiting $\alpha$ Centauri B with a semimajor axis $a_p = 1.5$ au with respect to the initial eccentricity $e_{po}$ and relative longitude $\Delta \varpi$ to the binary orbit.  Panel (a) illustrates numerical results using \texttt{Rebound}, while panel (b) shows the results from perturbation theory using $\epsilon_M$ \citep{Marchal1990} for the forced eccentricity.  Figure adapted from \citep{Quarles2018a}.}  
    \label{fig:ew_plane}
\end{figure}

However, the forced eccentricity is only one component, where we want to know the maximum eccentricity for which the planetary orbit can evolve. Equation \ref{eqn:forced_free} provides the components of the free eccentricity, where under secular evolution $e_{\rm max} = e_{\rm free} + e_F$ through vector addition. Figure \ref{fig:ew_plane} illustrates the differences between the two methods in the $(\Delta \varpi, e_{po})$ plane using the estimated maximum eccentricity (color-coded), where Fig. \ref{fig:ew_plane}a is obtained through n-body simulation, and Fig. \ref{fig:ew_plane}b is found using secular theory.  Comparing Fig. \ref{fig:ew_plane}a to Fig. \ref{fig:ew_plane}b, we see a fixed point that represents the forced eccentricity at the expected location.  The contours in Fig. \ref{fig:ew_plane}a at high values of $e_{po}$ do not rise as high as those in Fig. \ref{fig:ew_plane}b.  This shows that secular approximation falls short in capturing the a complete picture for $e_{po}>0.1$ even with a more accurate estimate of the forced eccentricity.

\begin{figure}
    \centering
    \includegraphics[width=0.75\linewidth]{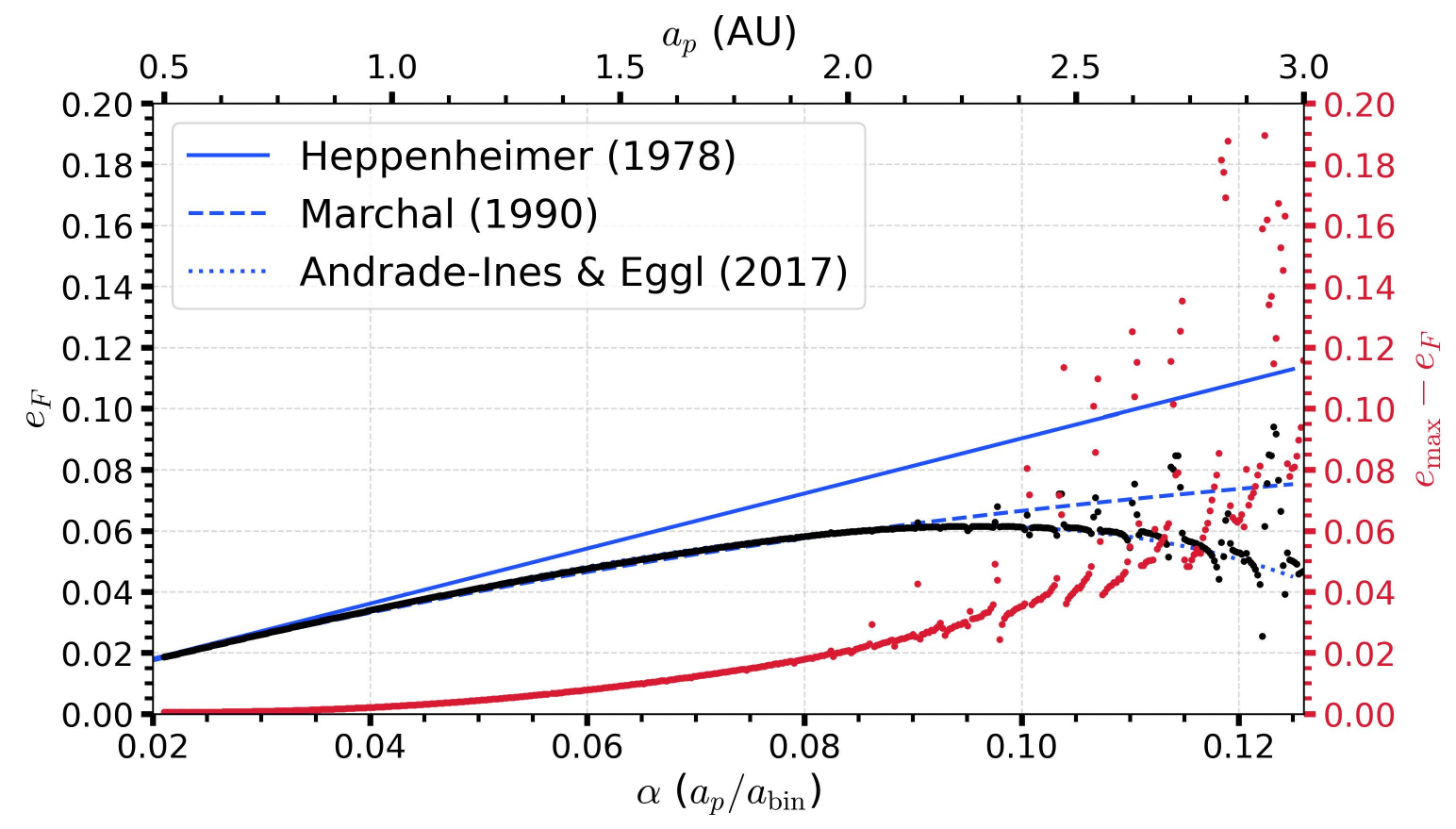}
    \caption{Forced eccentricity (black dots) for an {\bf S-Type} exoplanet orbiting $\alpha$ Centauri B $(\mu = 1.165)$ as a function of the initial semimajor ratio $\alpha$.  The approximation from \cite{Heppenheimer1978} overestimates $e_F$, where the approximation from \cite{Marchal1990} fits well for $\alpha \lesssim 0.09$.  The approximation from \cite{AndradeInes2017} captures the behavior for $\alpha >0.09$.  Data taken from \citep{Quarles2018a}.}
    \label{fig:eF_models}
\end{figure}

Using n-body simulations that minimize the eccentricity variation, we can estimate the forced eccentricity $e_F$ for a wide range in $\alpha$, where \cite{Quarles2018a} performed such simulations for a hypothetical planet in $\alpha$ Centauri AB.  Figure \ref{fig:eF_models} illustrates how the forced eccentricity determined by the n-body simulations (black dots) varies with $\alpha$, with an approximately quadratic growth.  The goodness of the analytical approaches depends on how well the solid lines match the dots. From Eq. \ref{eqn:forced_free}, we expect that the difference $e_{\rm max} - e_F$ is constant as both values would cancel each other to minimize the total eccentricity variation.  However, Fig. \ref{fig:eF_models} shows a growth in $e_{\rm max} - e_F$ (red dots), which demonstrates why the overall heights differed in Figure \ref{fig:ew_plane}a and \ref{fig:ew_plane}b.  An additional ${\sim}0.0125$ in $e_{\rm max}$ is unaccounted for in the secular approximation.  

This magnitude of $e_{\rm max} - e_F$ increases as $\alpha$ increases, but we can compare the approaches to estimate $e_F$ as that is independent of $e_{p}$ entirely.  The blue curves in Fig. \ref{fig:eF_models} show the predictions from each secular model, where all three converge as long as $\alpha \lesssim 0.035$.  Unsurprisingly, the \citeauthor{Heppenheimer1978} model diverges first and overestimates $e_F$.  The \cite{Marchal1990} model is accurate up to $\alpha \sim 0.09$ (or $a_p \sim 2.125$, where the \cite{AndradeInes2017} model most accurately predicts the full range.  Recall that the \cite{AndradeInes2017} model used a combination of numerical simulation and fitting techniques to determine its coefficients.  Overall, this analysis shows that a planet bound to $\alpha$ Cen B could stably orbit as long as $\alpha \lesssim 0.105$ (or $a_p \lesssim 2.5$ au), although its orbit is likely eccentric $(e_p \sim 0.1)$ due to the stellar companion.

\begin{figure}
    \centering
    \includegraphics[width=0.85\linewidth]{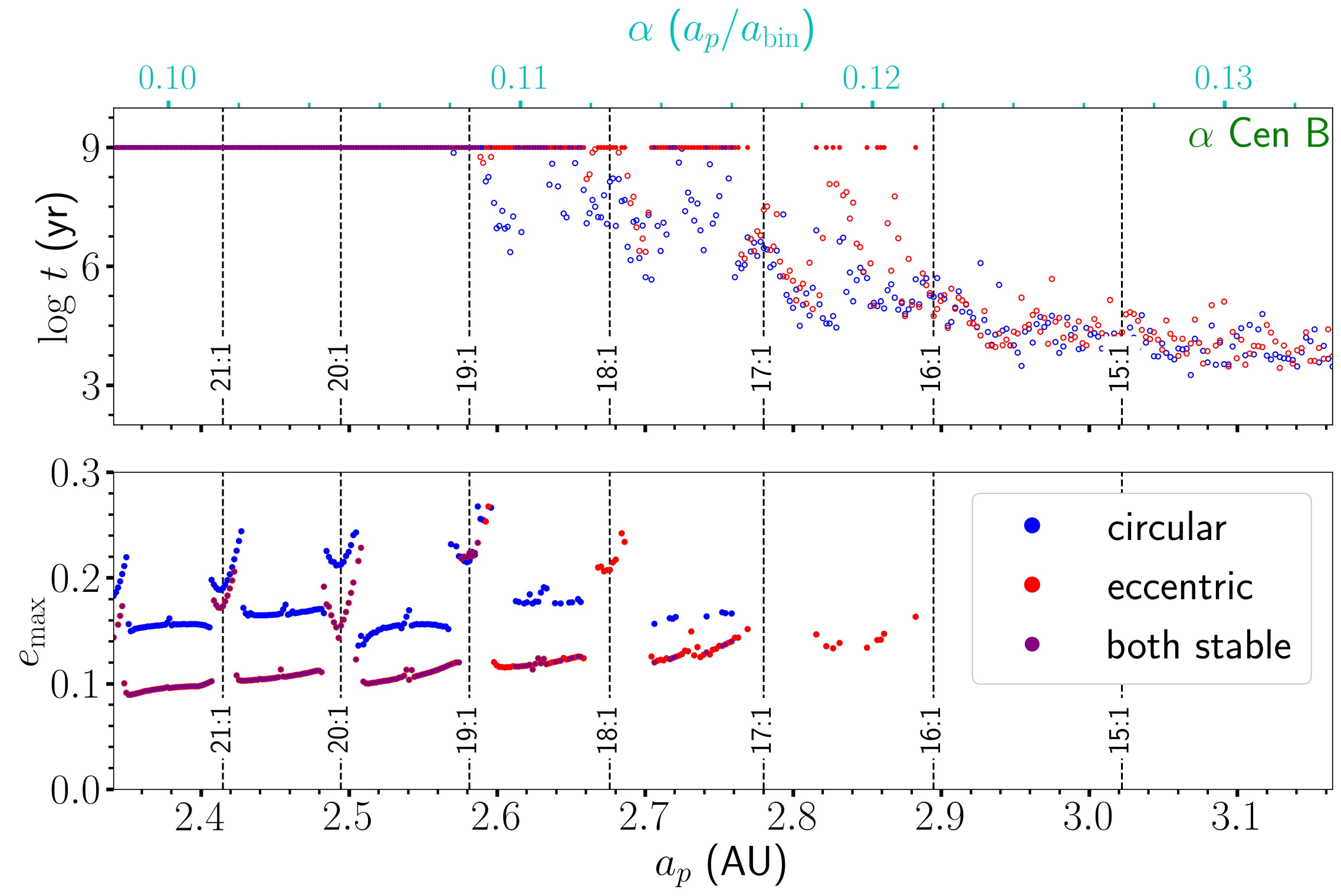}
    \caption{System lifetime $(\log\ t$; upper panel) and maximum eccentricity $(e_{\rm max}$; lower panel) for an {\bf S-Type} exoplanet orbiting $\alpha$ Cen B as a function of the initial planetary semimajor axis $a_p$ through long-term (1 Gyr) n-body simulations.  The exoplanet begins on either a circular orbit (blue dots) or at the forced eccentricity (red dots).  Vertical lines denote the approximate locations for $N:1$ mean motion resonances.  The top-axis provides a scale in terms of the semimajor axis ratio $\alpha$.  Figure adapted from \citep{Quarles2018a}.}
    \label{fig:aCenB_lifetime}
\end{figure}

A stability transition region (sometimes referred to as grey) appears as the forcing from the stellar companion increases (i.e., large $\alpha$).  \cite{Rabl1988} introduced the boundaries of this grey region as the upper critical orbit (UCO) or lower critical orbit (LCO) using numerical simulations for only 300 orbits of the stellar binary.  The transition from stable to unstable orbits is complicated by the choice for the initial planetary eccentricity, where either $e_{po} = 0$ (circular) or $e_{po} = e_F$ (eccentric) appear as good choices.  Figure \ref{fig:aCenB_lifetime} demonstrates the differences in the grey region's extent through billion-year numerical simulations of a hypothetical exoplanet orbiting $\alpha$ Cen B.  The maximum eccentricity for the planetary orbit is minimized for the initially eccentric (red dots) planetary orbits, while initially circular (blue dots) planetary orbits reach values about $50\%$ larger and even more for regions near $N:1$ mean motion resonances.  Note that our earlier estimate (based on Fig. \ref{fig:eF_models}) for the beginning of the grey region (or stability transition boundary) was within ${\sim}5\%$ of the value obtained from 1 Gyr n-body simulations.

\subsection{Stability through dynamical maps} \label{sec:dynmaps_Stype}
Secular approximations for the stability of an exoplanet on a coplanar orbit around either component in a stellar binary has its limitations.  Section \ref{sec:Stype_ecc} showed that the limitations become evident for large $(>0.1)$ values of $\alpha$, but approximations based on Eq. \ref{eqn:forced_free_params} can break down for circular binaries $(e_{\rm bin} = 0)$.  There are other tools from the circular restricted three-body problem (CRTBP) that help guide our inquiry into the planetary stability on S-Type orbits.  The Jacobi constant (or the integral of relative energy) parametrizes the transition from stable to unstable orbits.

\cite{Eberle2008} developed a stability criterion in the CRTBP using the Jacobi constant $C_J$ and semimajor axis ratio $\alpha$.   
In the restricted problem, the planet does not affect the orbital evolution of the stellar binary ($m_2 \rightarrow 0$), where it becomes more natural to redefine the binary mass ratio $\mu$ as

\begin{align}
\mu^\dagger &= \frac{m_3}{m_1 + m_3}.
\end{align}

In barycentric coordinates, $\mu^\dagger$ represents the fractional part of the binary semimajor axis for which $m_1$ orbits the common center-of-mass $(a_1 = \mu^\dagger a_{\rm bin})$.  It also provides a more constrained definition for the mass ratio when the exoplanet orbits the other star $(0\leq\mu^\dagger \leq1)$.  When $\mu^\dagger < 0.5$, the planet orbits $m_1$ and when $\mu^\dagger>0.5$, the masses $m_1$ and $m_3$ switch positions (assuming that $m_1 > m_3$).  The Jacobi constant \citep{Eberle2008} is then computed as

\begin{align}
C_J = \mu^\dagger + 2\mu^\dagger \alpha + \frac{1-\mu^\dagger}{\alpha} + \frac{2\mu^\dagger}{1+\alpha} + 2\sqrt{\alpha(1-\mu^\dagger)},
\end{align}

which depends on $\mu^\dagger$ and the initial semimajor axis ratio $\alpha$, but not on any orbital parameter attained over the system's evolution.  The Jacobi constant $C_J$ also parametrizes the zero-velocity contour, which bounds the exoplanet trajectories absent any close approaches.  Using the approximations for the collinear Lagrange equilibrium points, the Jacobi constant can be calculated for a particle at $L_1$, $L_2$, and $L_3$.  Starting an exoplanet using $L_1$ correspondingly produces a $C_1$, which should limit the planetary stability because $L_1$ corresponds to a unstable point in the gravitational potential.

\begin{figure}
    \centering
    \includegraphics[width=0.85\linewidth]{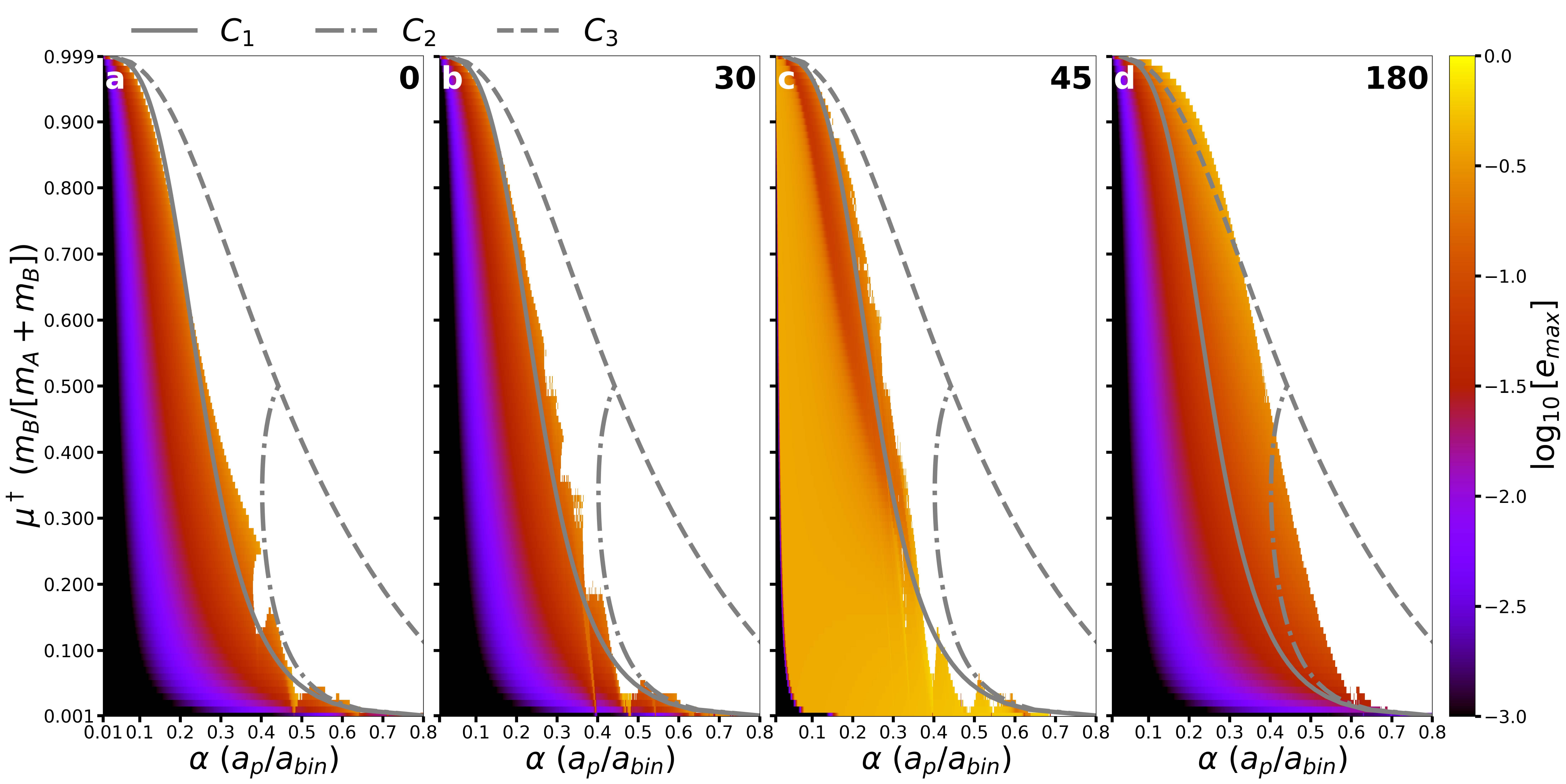}
    \caption{Stable configurations for an S-Type in the CRTBP (color-coded by $\log_{10}[e_{\rm max}$) with respect to the initial semimajor axis ratio $\alpha$ and mass ratio $\mu^\dagger$.  Four different initial planetary inclinations $(i_p = 0^\circ,\ 30^\circ,\ 45^\circ,$ and $180^\circ)$ are considered in panels a-d, where the unstable configurations are denoted by white cells.  The gray contours correspond to the initial parameters that place the putative planet at one of the collinear Lagrange points $(L_1,\ L_2,\ L_3)$.  Figure adapted from \citep{Quarles2020}.}
    \label{fig:CRTBP_max_ecc}
\end{figure}

Numerical n-body simulations \citep{Quarles2020} reveal that this is approximately true for inclined, prograde orbits.  Jacobi constants at the other two collinear points $(L_2$ and $L_3)$ provide more stringent limits in the $(\alpha,\mu^\dagger)$ parameter space.  Retrograde orbits $(i_p > 90^\circ)$ are not immune from this criterion, where $C_3$ bounds the stability the most.  Figure \ref{fig:CRTBP_max_ecc} illustrates the stability of an exoplanet in the CRTBP for planetary inclinations of $0^\circ$, $30^\circ$, $45^\circ$, and $180^\circ$ relative to the binary orbital plane.  The white regions are unstable, where the colored regions represent initial conditions that survive for at least $5 \times 10^5$ yr. 

Each panel is color-coded (on a logarithmic scale) with the maximum eccentricity attained over the full duration of system evolution.  The orbits of low inclination planets (Figs. \ref{fig:CRTBP_max_ecc}a and \ref{fig:CRTBP_max_ecc}b) have maximum eccentricity $e_{\rm max}$ values that increase with $\alpha$, but are very much limited by the inner Lagrange point defined by the $C_1$ curve.  Figure \ref{fig:CRTBP_max_ecc}c displays a similar structure, but the orbits have a much larger maximum eccentricity (due to the octopole term in the Hamiltonian; \cite{vonZeipel1910,Lidov1962,Kozai1962}) and is mostly independent of $\mu^\dagger$.  Figure \ref{fig:CRTBP_max_ecc}d shows that a retrograde orbiting exoplanet (i.e., orbiting opposite the direction of the stellar binary) can encompass much more of the $(\alpha,\mu^\dagger)$ parameter space, but is still mostly bounded by the $C_3$ curve.  The greater extent of stable retrograde orbits stems from a smaller average impulse imparted to the exoplanet on each passage of the stellar companion \citep[e.g.,][]{Hamilton1991}.

\begin{figure}
    \centering
    \includegraphics[width=0.45\linewidth]{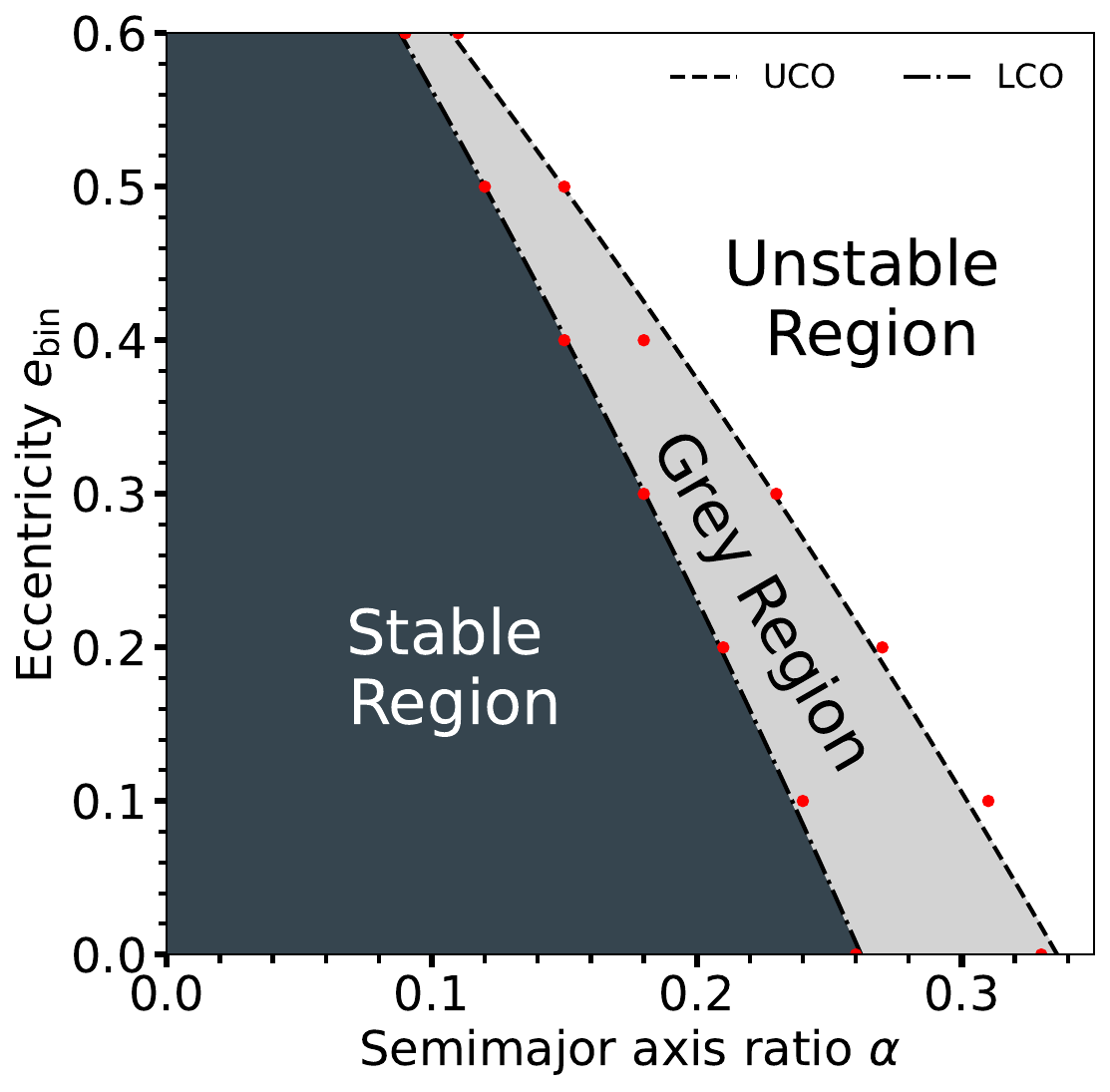}
    \caption{Stability regions for an S-Type exoplanet orbiting a single star in an equal-mass binary $(\mu^\dagger = 0.5)$ with respect to the initial semimajor axis ratio $\alpha$ and binary eccentricity $e_{\rm bin}$.  Red dots mark the critical semimajor axis ratio $\alpha_c$ using n-body simulations over 300 binary orbits and curves mark the least-squares fit (Eq. \ref{eqn:Rabl_stab})  determined by \cite{Rabl1988}.   Figure adapted from \citep{Rabl1988}.}
    \label{fig:Rabl_stab}
\end{figure}

Using n-body simulations, \cite{Rabl1988} identified stability regions considering exoplanets on S-Type orbits within {\it eccentric} binaries.  Note that others \citep[e.g.,][]{Chirikov1979} 
worked on quasi-analytic approaches to orbital stability based on Lyapunov characteristic numbers \citep{Lyapunov1892} and chaos theory. 
 The approach of \citeauthor{Rabl1988} developed a stability criterion for equal-mass binaries $(\mu^\dagger = 0.5,$ also known as the Copenhagen problem) that depended on the semimajor axis $a_p$ of the putative exoplanet and the binary eccentricity $e_{\rm bin}$.  They used relatively short (300 binary orbits) n-body simulations, but included eight equally spaced values in initial planetary longitude $\lambda_p$ and considered conditions where the binary began at its apastron or periastron position.   From their results, \citeauthor{Rabl1988} performed a least-squares quadratic fit to obtain a critical semimajor axis ratio $\alpha_c\ (= a_c/a_{\rm bin})$,

 \begin{align} \label{eqn:Rabl_stab}
 \alpha_c &= \begin{cases} (0.262 \pm 0.006) - (0.254 \pm 0.017)e_{\rm bin} - (0.060 \pm 0.027)e_{\rm bin}^2, & \text{(LCO)} \\
 (0.336 \pm 0.020) - (0.332 \pm 0.051)e_{\rm bin} - (0.082 \pm 0.082)e_{\rm bin}^2, & \text{(UCO)}
\end{cases}
\end{align}

for the lower critical orbit (LCO) and upper critical orbit (UCO), respectively.  Note that the errors in the fitted coefficients can be very large, especially in the quadratic term $e_{\rm bin}^2$.  Figure \ref{fig:Rabl_stab} illustrates the $(\alpha, e_{\rm bin})$ plane with the stability regions determined by \cite{Rabl1988}.  The stable region represents initial conditions that are independent of the planet's initial longitude $\lambda_p$ (i.e., all 8 trials survive the full simulation), where in the grey region some initial longitudes allow a planet to survive and other choices lead to escape.  Figure \ref{fig:Rabl_stab} provides a reasonable estimate for the orbital stability, but it is limited to a single binary mass ratio and a modest range in binary eccentricity.  These limitations reflect the computational capabilities of the time, where updates to the stability limit are possible proportionally to the advent of more efficient algorithms for orbital evolution and the greater availability of computing power.

The stability criterion for exoplanets on S-Type orbits was revisited by \cite{Holman1999}, who expanded the approach to include a finer grid in trial values for the semimajor axis ratio $(0.02 \leq \alpha \leq 0.50;$ with $\delta \alpha = 0.01)$ and a range in the binary mass ratio $(0.1\leq \mu^\dagger \leq 0.9;$ with $\Delta \mu^\dagger = 0.1)$.  Additionally, the duration of a simulation by \citeauthor{Holman1999} was increased from 300 to $10^4$ binary orbits.  This revision to the stability limit was possible due to the application of symplectic integration \citep{Wisdom1992} to n-body simulations and a huge expansion of computing power in personal computers.  \citeauthor{Holman1999} used eight equally-spaced values in the planetary longitude $\lambda_p$ following the procedure from \cite{Rabl1988}.  Incorporating the additional parameter in the binary mass ratio, the revised formula for the stability limit became

\begin{align} \label{eqn:Hol_stab_Stype}
\alpha_c =&\ (0.464 \pm 0.006) + (-0.380 \pm 0.010)\mu^\dagger + (-0.631 \pm 0.034)e_{\rm bin} + \nonumber \\
&\ (0.586 \pm 0.061)\mu^\dagger e_{\rm bin} + (0.150 \pm 0.041)e_{\rm bin}^2 + (-0.198 \pm 0.074)\mu^\dagger e_{\rm bin}^2.
\end{align}
Due to the relative simplicity of Eq. \ref{eqn:Hol_stab_Stype}, it became widely used (for 20 years) as the standard for determining the orbital stability of exoplanets on S-Type orbits.  Equation \ref{eqn:Hol_stab_Stype} is an empirical (not \emph{analytical}) formula that is determined through many n-body simulations that make certain assumptions.  Some of these assumptions are:

\begin{enumerate}
\item only three bodies exist in the system,
\item the simulation time $(10^4$ binary orbits) is sufficiently long,
\item the LCO boundary can be treated as the ``stability limit'',
\item the mass $m_2$ can be treated as a test particle (i.e., $m_2 \rightarrow 0)$,
\item all three bodies initially lie in the same plane (i.e., coplanar),
\item other effects on the planet's motion (e.g., tides or general relativity) are negligible.
\end{enumerate}

The known systems of binary stars with planets on S-Type orbits have not yet (largely) violated these assumptions, which has allowed the approach to remain in wide use.

\cite{Quarles2020} investigated how assumptions 2 and 5 affect the stability limit, as many of the confirmed planets on S-Type orbits do not lie in the stellar binary's orbital plane.   They also probe to finer ranges in $\mu^\dagger$, $e_{\rm bin}$, and $\alpha$, which allows for a revision to the formula for the stability limit through a similar least-squares procedure as \cite{Holman1999} and \cite{Rabl1988}.  For the coplanar case, the new stability formula is

\begin{align} \label{eqn:Quar_stab}
\alpha_c =&\ (0.501 \pm 0.002) + (-0.435 \pm 0.003)\mu^\dagger + (-0.668 \pm 0.009)e_{\rm bin} + \nonumber \\ 
&\ (0.644 \pm 0.015)\mu^\dagger e_{\rm bin} + (0.152 \pm 0.011)e_{\rm bin}^2 + (-0.196 \pm 0.019)\mu^\dagger e_{\rm bin}^2
\end{align}

which largely revises the lower order terms in $\mu^\dagger$ and $e_{\rm bin}$ because \cite{Quarles2020} extended their range in mass ratio down to $\mu^\dagger = 0.001$.  \cite{Holman1999} and \cite{Quarles2020} find similar values for the stability limit in $\alpha$ Centauri, which can be compared to more detailed studies of the system \citep[e.g.,][]{Quarles2016}.  These previous investigations show that the stability limit is drastically different for highly inclined (relative to the binary orbital plane) planets. 

\begin{figure}[!t]
    \centering
    \includegraphics[width=0.85\linewidth]{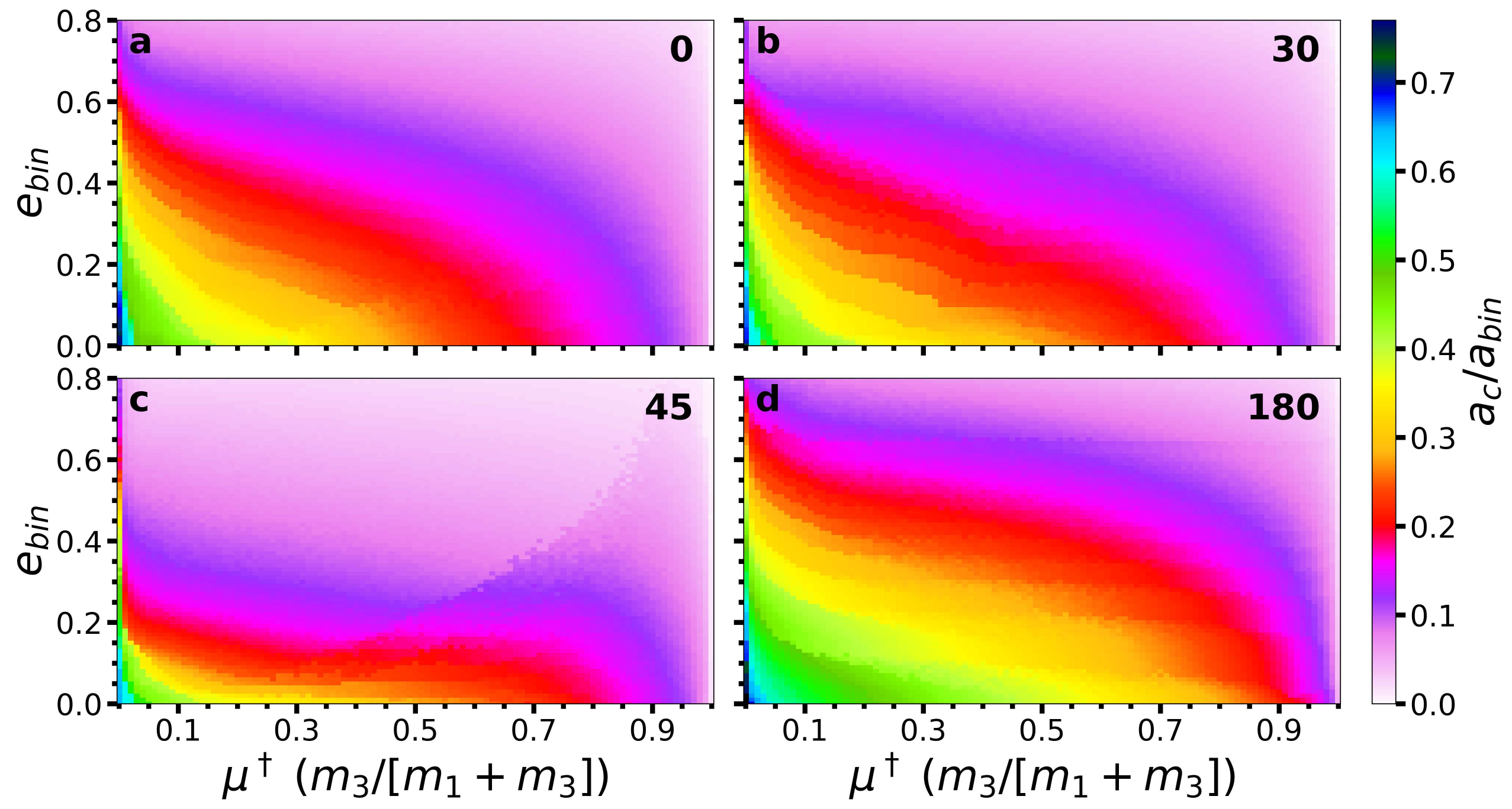}
    \caption{Critical semimajor axis $\alpha_c$ (color-coded) as a function of the binary eccentricity $e_{bin}$ and the mass ratio $\mu^\dagger$. The initial planetary inclination $i_p$ (in degrees) of the S-Type planet is denoted in each panel in the upper right.  Figure adapted from \citep{Quarles2020}.}
    \label{fig:Incl_stab}
\end{figure}

Figure \ref{fig:Incl_stab} illustrates how the critical semimajor axis $\alpha_c$ varies with the binary mass ratio $\mu^\dagger$ and eccentricity $e_{\rm bin}$.  For coplanar orbits (Fig. \ref{fig:Incl_stab}a), the contours for the critical semimajor axis $\alpha_c$ \emph{do not} appear to follow easily identifiable polynomials.  Therefore, we may expect either empirical formula (Eq. \ref{eqn:Hol_stab_Stype} or \ref{eqn:Quar_stab}) to fail near the edges of the map.  Figure \ref{fig:Incl_stab}b is similar to Fig. \ref{fig:Incl_stab}a with slight distortions in the contours for a given $\alpha_c$.  Figures \ref{fig:Incl_stab}c and \ref{fig:Incl_stab}d are even more distorted, where retrograde orbits (Fig. \ref{fig:Incl_stab}d) greatly enhance the extent of $\alpha_c$.  In contrast, sufficiently inclined orbits in Fig. \ref{fig:Incl_stab}c show that the von Zeipel-Lidov-Kozai mechanism \citep{vonZeipel1910,Lidov1962,Kozai1962}) is effective to excite the planetary orbit to very high values, which places an upper limit $\alpha{\sim}0.05$ in most cases.  

\cite{Quarles2018c} investigated the stability of multiplanet systems, specifically in $\alpha$ Centauri, using a similar formalism for planet packing.  Multiplanet systems need significantly more space between the planets due to the location of the single planet stability limit and the forced eccentricity from the stellar companion.

The process of optimizing a multivariate least-squares function tends to settle on coefficients that represent the median case (i.e., $\mu^\dagger{\sim}0.5$ and $e_{\rm bin}{\sim}0.5$).  Applications of the above empirical formulas towards high $(\mu^\dagger >0.9$) or low $(\mu^\dagger < 0.1)$ mass ratio give less accurate results when compared with n-body simulations specific to those ranges.  To avoid these issues, \cite{Quarles2020} provided a publicly available lookup table\footnote{see the GitHub repo: \href{https://github.com/saturnaxis/ThreeBody_Stability}{saturnaxis:ThreeBody\_Stability}}, where interpolation routines from \texttt{scipy.interpolate} can be implemented to provide accurate results and potentially go beyond the given grid resolution.  An example is shown as:

\begin{BoxTypeA}[chap1:box1]{Example 2D interpolation code in \texttt{python}}
\begin{lstlisting}[language=Python]
from scipy.interpolate import CloughTocher2DInterpolator
import numpy as np

mu = 0.05
e_bin = 0.5

Quarles_repo = "https://raw.githubusercontent.com/saturnaxis/ThreeBody_Stability/master/"
fn = "a_crit_Incl[0].txt"
X, Y, Z = np.genfromtxt(Quarles_repo + fn,delimiter=',',comments='#',unpack=True)
interp = CloughTocher2DInterpolator(np.array([X,Y]).T, Z)

print("a_c = ",interp(mu,e_bin))
\end{lstlisting}
\end{BoxTypeA}

\section{P-type stability}\label{sec:Ptype}
\begin{figure}[!b]
    \centering
    \includegraphics[width=\linewidth]{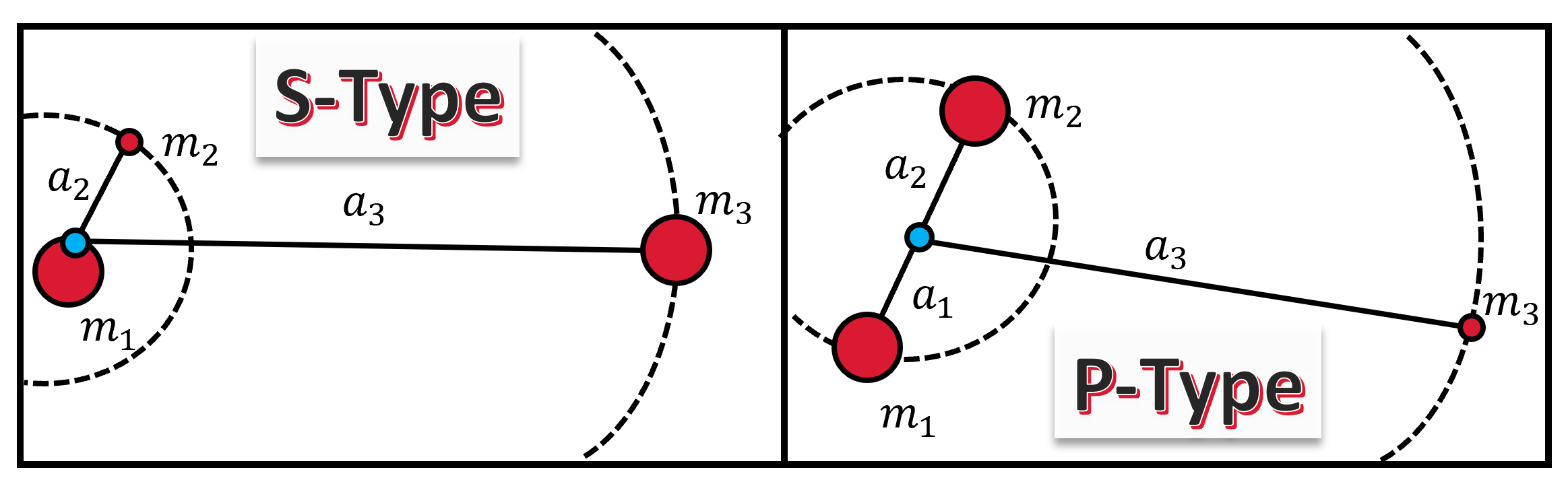}
    \caption{Schematic illustrating the difference between planets in either S-Type (left) or P-Type (right) orbits. The stellar masses are represented as the larger dots, where the planet is the smaller dot in each panel.}
    \label{fig:SType_PType}
\end{figure}

Planets on P-Type orbits may appear more familiar because such planets orbit both stars completely (see Fig. \ref{fig:SType_PType}).  This orbital architecture is actually similar to the S-Type orbits describe in Section \ref{sec:Stype}.  Let us consider Figure \ref{fig:arch}, but define $m_2=m_B$ (the stellar companion's mass) and $m_3 = m_p$ (the planetary mass).  Consequently, the semimajor axes are also redefined with $a_{\rm bin} = a_1 + a_2$ and $a_3 = a_p$. By doing so, we have effectively switched indices in the hierarchy and are able to re-use many of the same theoretical concepts and techniques.  

\subsection{Secular evolution through a disturbing function}
A planet's orbit can change secularly, assuming that the planet begins far from the inner region that is dominated by the N:1 mean motion resonances with the binary (i.e., $a_p \gg a_{\rm bin}$.  In this formulation, we also make the same assumptions that the planetary eccentricity and/or orientation evolve while variations in the semimajor axis are negligible (i.e., $\dot{a}_p \approx 0)$.  The secular interaction with the inner binary is modeled using a Laplace-Lagrange disturbing function, which can be expanded in terms of the semimajor axis ratio $(a_p/a_{\rm bin})$ using Legendre polynomials $(P_i)$.  In the quadrupole limit, \cite{Moriwaki2004} obtained the following secular disturbing function in orbital elements:

\begin{align} \label{eqn:P_disturb}
\mathcal{R}_{\rm MN} = \mu^\dagger(1-\mu^\dagger)n_{\rm bin}^2 a_{\rm bin}^2 \left(\frac{a_p}{a_{\rm bin}}\right)^{-3}\left[\frac{1}{4}\left(1+\frac{3}{2}e_p^2\right)\left(1+\frac{3}{2}e_{\rm bin}^2\right) - \frac{15}{16}\left(1-2\mu^\dagger \right) \left(\frac{a_p}{a_{\rm bin}}\right)^{-1} e_p \left( e_{\rm bin} + \frac{3}{4}e_{\rm bin}^3\right) \cos\Delta\varpi \right],
\end{align}

where the subscript $p$ and $\rm bin$ refer to the planetary and binary orbital elements, respectively.   Note that Eq. \ref{eqn:P_disturb} is truncated in a similar manner as Eq. {\ref{eqn:S_Marchal}} in the S-Type case. The relative longitude $\Delta \varpi\ (=\varpi_p - \varpi_{\rm bin})$ represents the relative orientation between the planetary and binary orbits within a reference plane.  The binary mass ratio $\mu^\dagger\ \left[=m_B/(m_A+m_B)\right]$ and mean motion $n_{\rm bin}\ \left(=\sqrt{G(m_A+m_B)/a_{\rm bin}^2} \right)$ are used as prefactors due to our switching of indices in the setup of the orbital architecture.  Note that this disturbing function has a similar structure as Eq. \ref{eqn:S_disturb}, where the higher order terms in $e_p$ and $e_{\rm bin}$ are ignored.

Using the eccentricity vector components in Eq. \ref{eqn:ecc_vec}, the $\cos(a-b)$ trigonometric identity, and the semimajor axis ratio $\alpha\ (= a_p/a_{\rm bin})$, the disturbing function (Eq. \ref{eqn:P_disturb}) is rewritten as

\begin{align}
\begin{aligned}
\mathcal{R}_{\rm MN} = \mu^\dagger(1-\mu^\dagger)n_{\rm bin}^2 a_{\rm bin}^2 
\alpha^{-3}\left[\frac{1}{4}\left(1+\frac{3}{2}h^2+\frac{3}{2}k^2\right)\left(1+\frac{3}{2}e_{\rm bin}^2\right) - \frac{15}{16}\left(1-2\mu^\dagger \right) \alpha^{-1} e_p \left( e_{\rm bin} + \frac{3}{4}e_{\rm bin}^3\right) \left(k\cos\varpi_{
\rm bin} - h\sin\varpi_{\rm bin}\right)\right],
\end{aligned}
\end{align}

so that we can apply Hamilton's equations (Eq. \ref{eqn:hamilton}) to get 

\begin{align}\label{eqn:sec_ecc_P}
\begin{aligned}
\dot{k} &= -g_{MN}h -B\sin\varpi_{\rm bin}, \\
\dot{h} &= g_{MN}k - B\cos\varpi_{\rm bin},
\end{aligned}
\end{align}
where

\begin{align}
g_{MN} &= \frac{3}{4}\mu^\dagger\left(1-\mu^\dagger\right) \alpha^{-5}\frac{n_{\rm bin}^2}{n_p}  \left(1+\frac{3}{2}e_{\rm bin}^2\right), \\
B &= \frac{5}{4}\left(1-2\mu^\dagger\right)g_{MN}\alpha^{-1} e_{\rm bin}\left[\frac{1+(3/4)e_{\rm bin}^2}{1+(3/2)e_{\rm bin}^2}\right].
\end{align}

Using the method to solve coupled first-order equations, we find equations that resemble a forced harmonic oscillator as:

\begin{align}
\begin{aligned}
\ddot{h} + g_{MN}^2h &= -g_{MN}B\sin\varpi_{\rm bin}, \\
\ddot{k} + g_{MN}^2k &= - g_{MN}B\cos\varpi_{\rm bin},
\end{aligned}
\end{align}
which has the analytical solutions:

\begin{align}
\begin{aligned}
k(t) &= e_{\rm free}\cos(g_{MN}t + \varphi) + \epsilon_{MN}\cos\varpi_{\rm bin}, \\
h(t) &= e_{\rm free}\sin(g_{MN}t + \varphi) + \epsilon_{MN}\sin\varpi_{\rm bin}
\end{aligned}
\end{align},
where $e_{\rm free}$ and $\varphi$ are determined by the initial conditions $k(0)$ and $h(0)$.  The forced eccentricity $\epsilon_{MN}$ is given by

\begin{align}\label{eqn:forced_MN}
\epsilon_{MN} &= \frac{B}{g_{MN}} = \frac{5}{4}\left(1-2\mu^\dagger\right)\alpha^{-1} e_{\rm bin}\left[\frac{1+(3/4)e_{\rm bin}^2}{1+(3/2)e_{\rm bin}^2}\right].
\end{align}

This secular approximation is applicable to a broad range of cases, but it clearly has limits.  For example, the symmetric cases $e_{\rm bin}$ (circular binary) or $\mu^\dagger = 0.5$ (equal mass binary) result in $\epsilon_{MN} = 0$.  N-body simulations of disks (using test particles) show that a forced eccentricity exists that pumps the eccentricity of disk particles \citep{Moriwaki2004}.  

\begin{figure}
    \centering
    \includegraphics[width=0.75\linewidth]{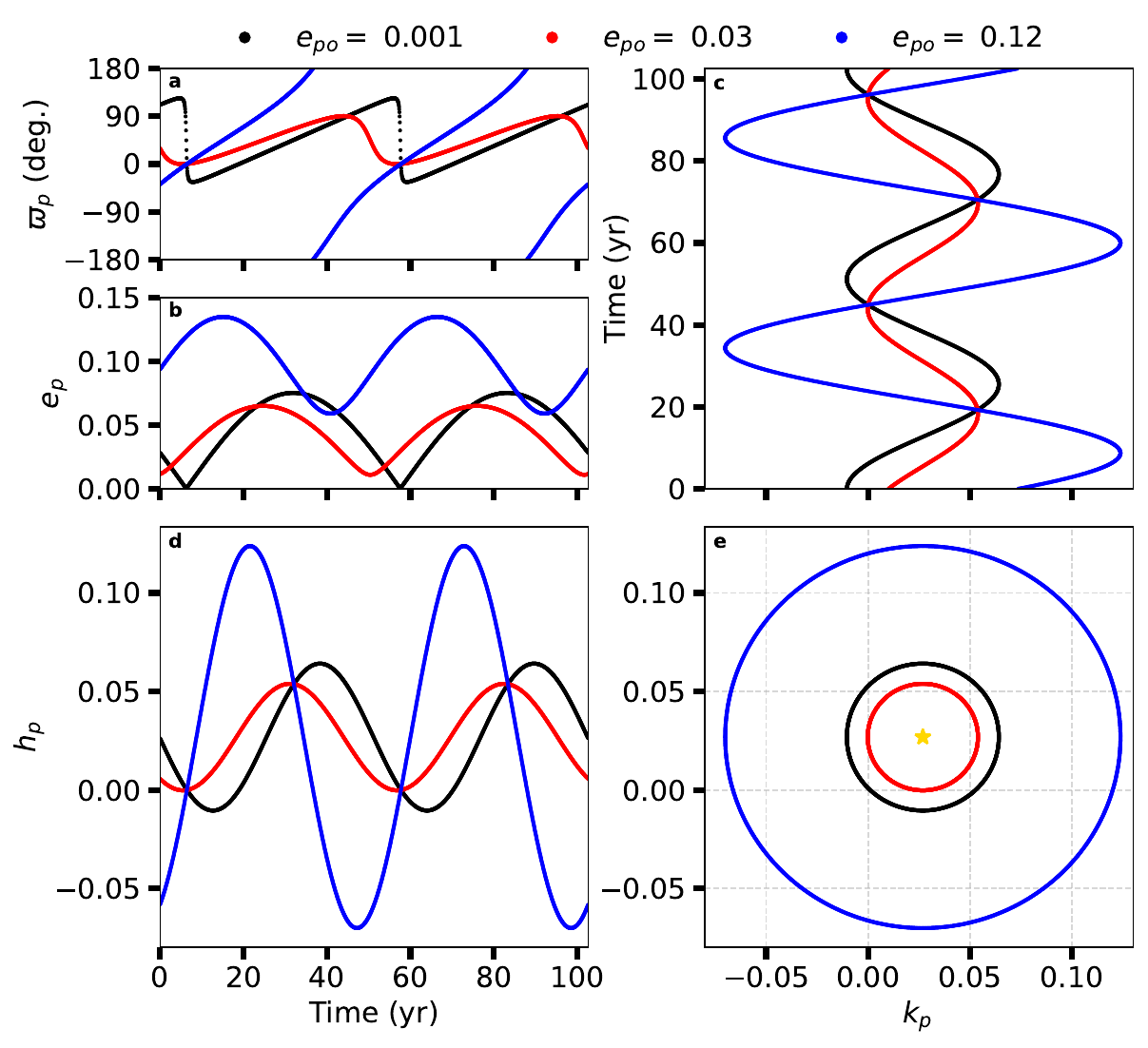}
    \caption{Secular time evolution for an P-Type planet using the \cite{Moriwaki2004} model, where $m_1=1\ M_\odot$, $m_2 = 0.3\ M_\odot$, $\alpha = 5$, $e_{\rm bin}=0.3$ for three different initial eccentricities $e_{po}$ initially misaligned with the binary orbit $(\Delta\varpi = \varpi_{\rm bin} = 45^\circ)$.  The gold star in panel e represents the forced eccentricity $\epsilon_{MN}$, which does not depend on $e_{po}$. }
    \label{fig:MN_sec}
\end{figure}

Figure \ref{fig:MN_sec} illustrates the secular evolution (assuming three initial eccentricities) of a hypothetical planet in a P-Type orbit, where $m_1 = 1\ M_\odot$, $m_2 = 0.3\ M_\odot$, $\alpha = 5$, and $e_{\rm bin }= 0.3$.  The evolution using the \citeauthor{Moriwaki2004} model appears similar to Fig. \ref{fig:Hep_sec}, however the planet is initially misaligned with the binary orbit $(\Delta \varpi = 45^\circ)$.  This offsets the evolution of the planetary longitude of pericenter $\varpi_p$ (Fig. \ref{fig:MN_sec}a), and the eccentricity vectors $k_p$ (Fig. \ref{fig:MN_sec}c) and $h_p$ (Fig. \ref{fig:MN_sec}d).  As a result, the center point (gold star in Fig. \ref{fig:MN_sec}e) is rotated (counter-clockwise) by $\varpi_{\rm bin}$.

\subsection{Applications of forced and free eccentricity} \label{sec:Ptype_ecc}

Similar to Sec. \ref{sec:Stype_ecc}, the forced and free eccentricity can provide a good indicator for the stability of an exoplanet on a P-Type orbit.  The forced eccentricity $\epsilon_{MN}$ depends on the shape of the stellar binary's orbit $(a_{\rm bin}$ and $e_{\rm bin})$ and the planetary semimajor axis $a_p$, since $\alpha = a_p/a_{\rm bin}$.  However, the forced eccentricity is expanded in inverse powers using the ratio $\alpha$ since $a_p \gg a_{\rm bin}$.  The forced eccentricity sets a reference point for which the planetary eccentricity oscillates around (see Fig. \ref{fig:MN_sec}).

\cite{Moriwaki2004}, \cite{Paardekooper2012}, and \cite{Demidova2015} used calculations of the forced eccentricity to better understand structures within circumbinary protoplanetary disks.  \cite{Paardekooper2012} investigated planet formation of the newly discovered ``Tatooine'' exoplanets from the Kepler Space telescope, in which they derive a fast component to describe a growing planet's eccentricity evolution by averaging over the mean longitude of the inner binary only (instead of averaging over both) in the disturbing function.  As a result, they obtained a \emph{fast} version of the forced eccentricity $\epsilon_{ff}$, where

\begin{align}
\epsilon_{ff} &= \frac{3}{4}\mu^\dagger\left(1-\mu^\dagger \right) \alpha^{-2}\sqrt{1+\frac{34}{3}e_{\rm bin}^2}.
\end{align}

Note that $\epsilon_{ff}$ falls off faster than $e_{MN}$ (Eq. \ref{eqn:forced_MN}) as $\alpha$ increases.  \cite{Demidova2015} used the forced eccentricity $\epsilon_{MN}$ to model eccentricity variations and spiral patterns of particles within the circumbinary protoplanetary disk, where $e_p = 2\epsilon_{MN}|\sin(g_{MN}t)|$.  Through this model, one can approximate the time-averaged eccentricity as a combination of the fast $\epsilon_{ff}$ and slow $\epsilon_{MN}$ components \citep{Shevchenko2018}, or

\begin{align}
\langle e_p \rangle = \frac{4}{\pi}\left(\epsilon_{MN} + \epsilon_{ff}\right).
\end{align}

\begin{figure}
    \centering
    \includegraphics[width=0.75\linewidth]{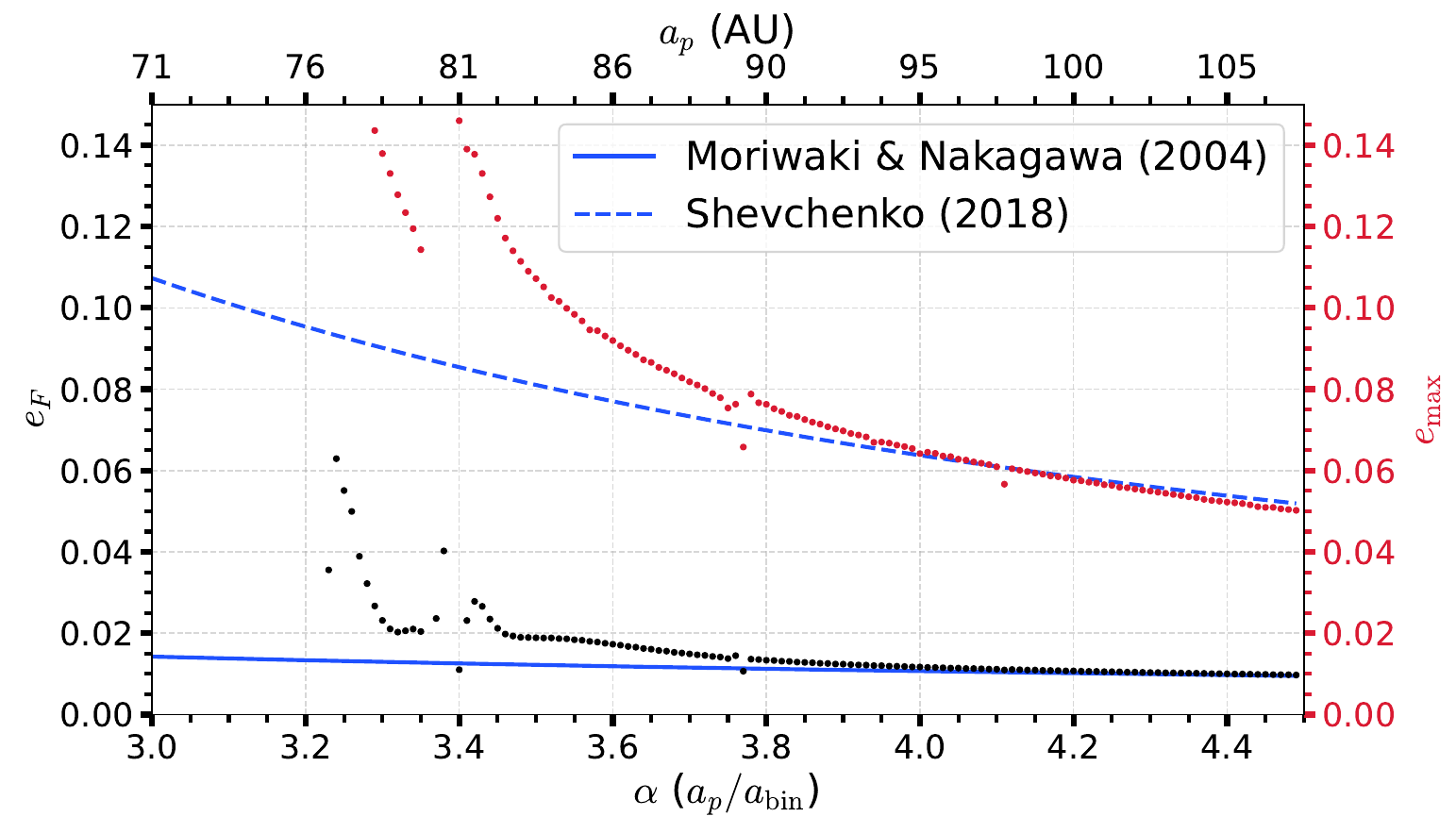}
    \caption{Forced eccentricity (black dots) for a {\bf P-Type} exoplanet orbiting $\alpha$ Centauri AB $(\mu^\dagger = 0.461)$ as a function of the initial semimajor ratio $\alpha$.  The approximation (Eq. \ref{eqn:forced_MN}) from \cite{Moriwaki2004} underestimates $e_F$ when $\alpha \lesssim 4$.  The approximation from \cite{Shevchenko2018} captures the behavior of $e_{\rm max}$ for $\alpha \gtrsim 4$.  Data taken from \citep{Quarles2018a}.}
    \label{fig:eF_models_Ptype}
\end{figure}

Using n-body simulations that minimize the eccentricity variation, we can estimate the forced eccentricity $e_F$ for a wide range in $\alpha$, where \cite{Quarles2018a} performed such simulations for a hypothetical \emph{circumbinary} planet in $\alpha$ Centauri AB $(\mu^\dagger = 0.461$, $e_{\rm bin} = 0.524$, and $a_{\rm bin} = 23.78\ {\rm au})$.  Figure \ref{fig:eF_models_Ptype} illustrates how the forced eccentricity (black dots) and maximum eccentricity (red dots) determined by n-body simulations varies with $\alpha$, where the magnitude of each decays with increasing $\alpha$.  The secular approximation from \cite{Moriwaki2004} (solid blue line) is accurate for $\alpha \gtrsim 4$, but underestimates forced eccentricity $e_F$ for planets closer to the inner binary due to additional effects (e.g., $N$:1 mean motion resonances).  The dashed blue line shows the approximation $\langle e_p \rangle$ using both approximations for the forced eccentricity, which closely approximates $e_{\rm max}$ for $\alpha \gtrsim 4$.  However, the ratio $e_{\rm max}/e_F$ converges to 5 rather than 2, as we would expect from the addition of the free and forced eccentricity vectors.

\subsection{Stability through dynamical maps}
Similar to S-Type systems, exoplanets on P-Type orbits also suffer limitations when using the secular approximation.  In particular, Eq. \ref{eqn:forced_MN} goes to zero when the binary orbit is circular or for equal-mass binaries (i.e., $\mu^\dagger = 0.5)$.  Other works have provided a more detailed expansion in which other forcing terms are included.  These expansions have informed the studies of circumbinary planets \citep{Li2014} and the evolution of Pluto's smaller moons \citep{Bromley2015b}.  However, these other terms in the expansion depend on factors of

\begin{align*}
\frac{m_1-m_2}{m_1+m_2} &= 1-2 \mu^\dagger,
\end{align*}

which goes to zero, when $\mu^\dagger = 0.5$.  Equal-mass binaries appear symmetric in the secular potential, where $e_F \approx 0$ makes sense.  But, other (non-secular) factors persist to perturb a circumbinary planet so that its orbit is highly eccentric (see Fig. \ref{fig:eF_models_Ptype} at $\alpha = 3.2-3.4$).  

To explore all possible cases, we again turn to n-body simulations and how those numerical solutions have helped to provided a more general criterion for orbital stability.  \cite{Dvorak1986} developed a stability criterion that for equal-mass binaries $(\mu^\dagger = 0.5)$ that depended only on the binary eccentricity $e_{\rm bin}$ and planetary semimajor axis $a_p$ to produce the critical semimajor axis ratio $\alpha\ (=a_p/a_{\rm bin})$.  At the time and sometimes even now, an exact definition of orbital stability is not always given.  It is important to note that \citeauthor{Dvorak1986} contended with two definitions:

\begin{enumerate}
    \item "A stable orbit is defined as an orbit having elliptic orbital elements with an eccentricity smaller than 0.3 throughout the whole integration time of 500 periods of the primary bodies."
    \item "All solutions stay in bounded regions of the phase space and no collisions and no escapes of the bodies occur."
\end{enumerate}

The first definition is more precise, but the conditions given (e.g., $\text{max}\ e_p < 0.3$) were defined relative to numerical experience.  But, the perturbations act in a similar way on the orbital elements of stable and unstable trajectories alike for the "escape" orbits within the phase space so that the second definition is also found wanting.  We include this dilemma to show the importance of including your assumptions and that trade-offs are often necessary with limited resources.

\begin{figure}
    \centering
    \includegraphics[width=0.45\linewidth]{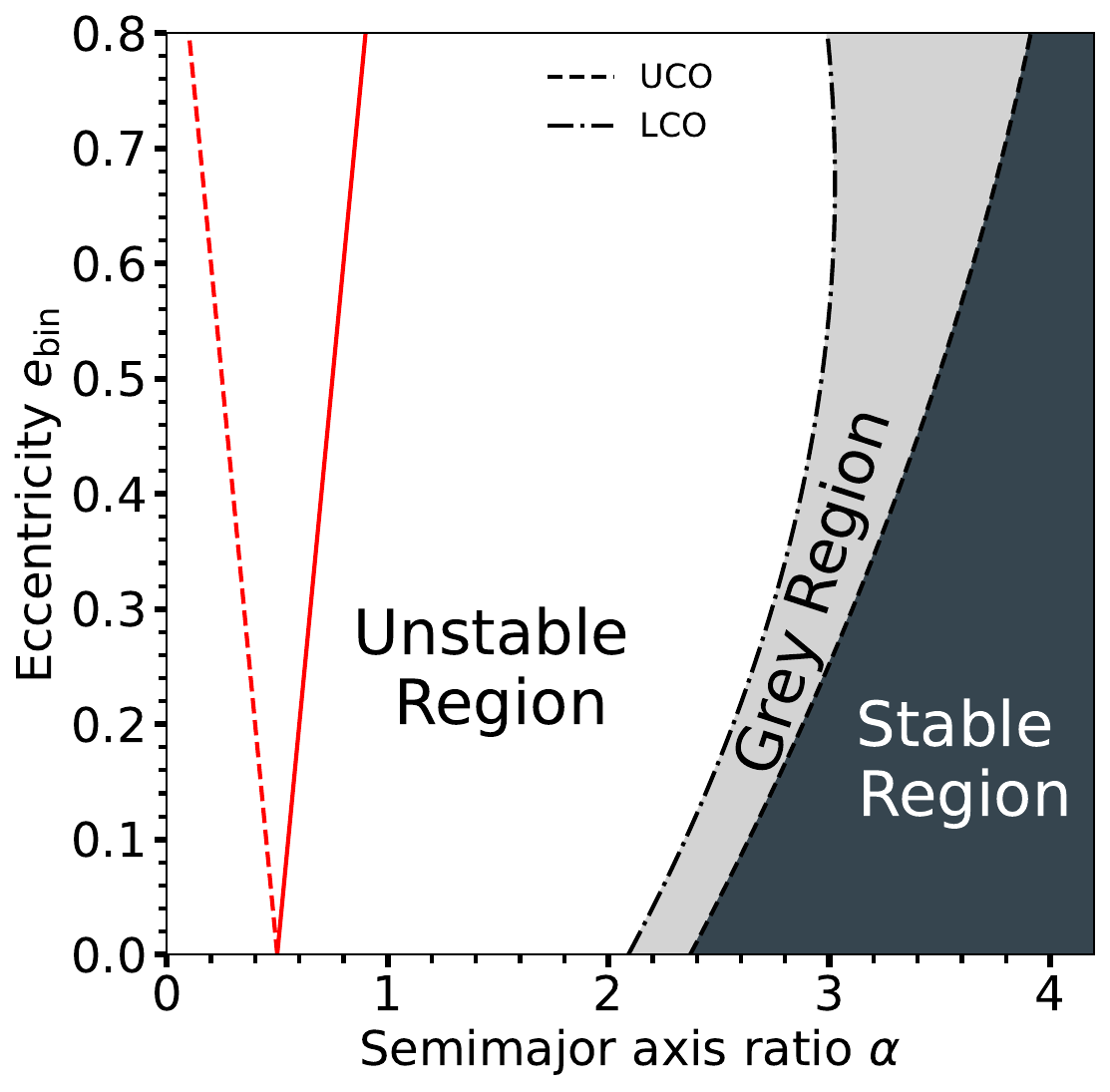}
    \caption{Stability regions for  equal-mass binary $(\mu^\dagger = 0.5)$ with respect to the initial semimajor axis ratio $\alpha$ and binary eccentricity $e_{\rm bin}$.  The dashed and dash-dot curves mark the least-squares fit (Eq. \ref{eqn:Dvorak_stab})  determined by \cite{Dvorak1986}. The red curves illustrate the relative proximity (relative to the binary semimajor axis) of the secondary star at apastron (solid) or periastron (dashed).  Figure adapted from \cite{Dvorak1986}.}
    \label{fig:Dvorak_stab}
\end{figure}

\citeauthor{Dvorak1986} used short (400 binary orbits) n-body simulations, but included ten equally-spaced values in the initial planetary longitude $\lambda_p$ and considered conditions where the binary began at its apastron or periastron position; the approach in \cite{Rabl1988} was based on this technique.  From his results, \citeauthor{Dvorak1986} used a least-squares quadratic fit to obtain two relations for the critical semimajor axis,

 \begin{align} \label{eqn:Dvorak_stab}
 \alpha_c &= \begin{cases} (2.09 \pm 0.30) + (2.79 \pm 0.53)e_{\rm bin} - (2.08 \pm 0.56)e_{\rm bin}^2, & \text{(LCO)} \\
 (2.37 \pm 0.23) + (2.76 \pm 0.40)e_{\rm bin} - (1.04 \pm 0.43)e_{\rm bin}^2, & \text{(UCO)}
\end{cases}
\end{align}

for the lower critical orbit (LCO) and upper critical orbit (UCO), respectively.  Figure \ref{fig:Dvorak_stab} shows the $(\alpha,\ e_{\rm bin})$ plane with the stability regions determined by \cite{Dvorak1986}.  The stable region represents where initial conditions are independent of the planet's initial longitude $\lambda_p$ (i.e., all 10 trials survive the full simulation), where in the grey region some initial longitudes allow a planet to survive while others initial values lead to escape.

Similar to S-Type orbits, the stability criterion for exoplanets on P-Type orbits was revisited by \cite{Holman1999}, who expanded the approach to include a finer grid in trial values for the semimajor axis ratio $(1 \leq \alpha \leq 5;$ with $\delta \alpha = 0.1)$ and a range in the binary mass ratio $(0.1\leq \mu^\dagger \leq 0.5;$ with $\Delta \mu^\dagger = 0.1)$.  Additionally, the duration of a simulation by \citeauthor{Holman1999} was increased from 400 to $10^4$ binary orbits due to the application of symplectic integration \citep{Wisdom1992} to n-body simulations and a huge expansion of computing power in personal computers.  \citeauthor{Holman1999} used eight equally-spaced values in the planetary longitude $\lambda_p$ following the procedure from \cite{Dvorak1986}.  Incorporating the additional parameter in the binary mass ratio, the revised formula for the stability limit became

\begin{align} \label{eqn:Hol_stab_Ptype}
\alpha_c =&\ (1.60 \pm 0.04) + (4.12 \pm 0.09)\mu^\dagger + (5.10 \pm 0.05)e_{\rm bin} + (-4.27 \pm 0.17)\mu^\dagger e_{\rm bin} +  \nonumber \\ &\ (-2.22 \pm 0.11)e_{\rm bin}^2 + (-5.09 \pm 0.11)\left(\mu^\dagger\right)^2 + (4.61 \pm 0.36)\left(\mu^\dagger e_{\rm bin} \right)^2.
\end{align}

Due to the relative simplicity of Eq. \ref{eqn:Hol_stab_Ptype}, it became widely used (for 20 years) as the standard for determining the orbital stability of exoplanets on P-Type orbits.  Equation \ref{eqn:Hol_stab_Ptype} is an empirical (not \emph{analytical}) formula that is determined through many n-body simulations that make certain assumptions.  Most of the same assumptions outlined in Sec. \ref{sec:dynmaps_Stype} apply, except assumption \#3 uses the UCO boundary instead.

\cite{Quarles2018b} investigated how assumptions $1-4$ affect our estimate of the stability limit and how the newly discovered circumbinary planets span the $(\mu^\dagger,\ e_{\rm bin})$ plane for orbital stability.  Note that Kepler-47 is a 3 planet circumbinary system \citep{Orosz2019}, where only 2 planets were known at the time of initial discovery \citep{Orosz2012}.  The 2 planet solution to Kepler-47 motivated \cite{Quarles2018b} to use methods developed in the chapter for compact systems orbiting single stars to account for the outer planet's influence on the orbital stability of the inner planet.

\begin{figure}
    \centering
    \includegraphics[width=0.85\linewidth]{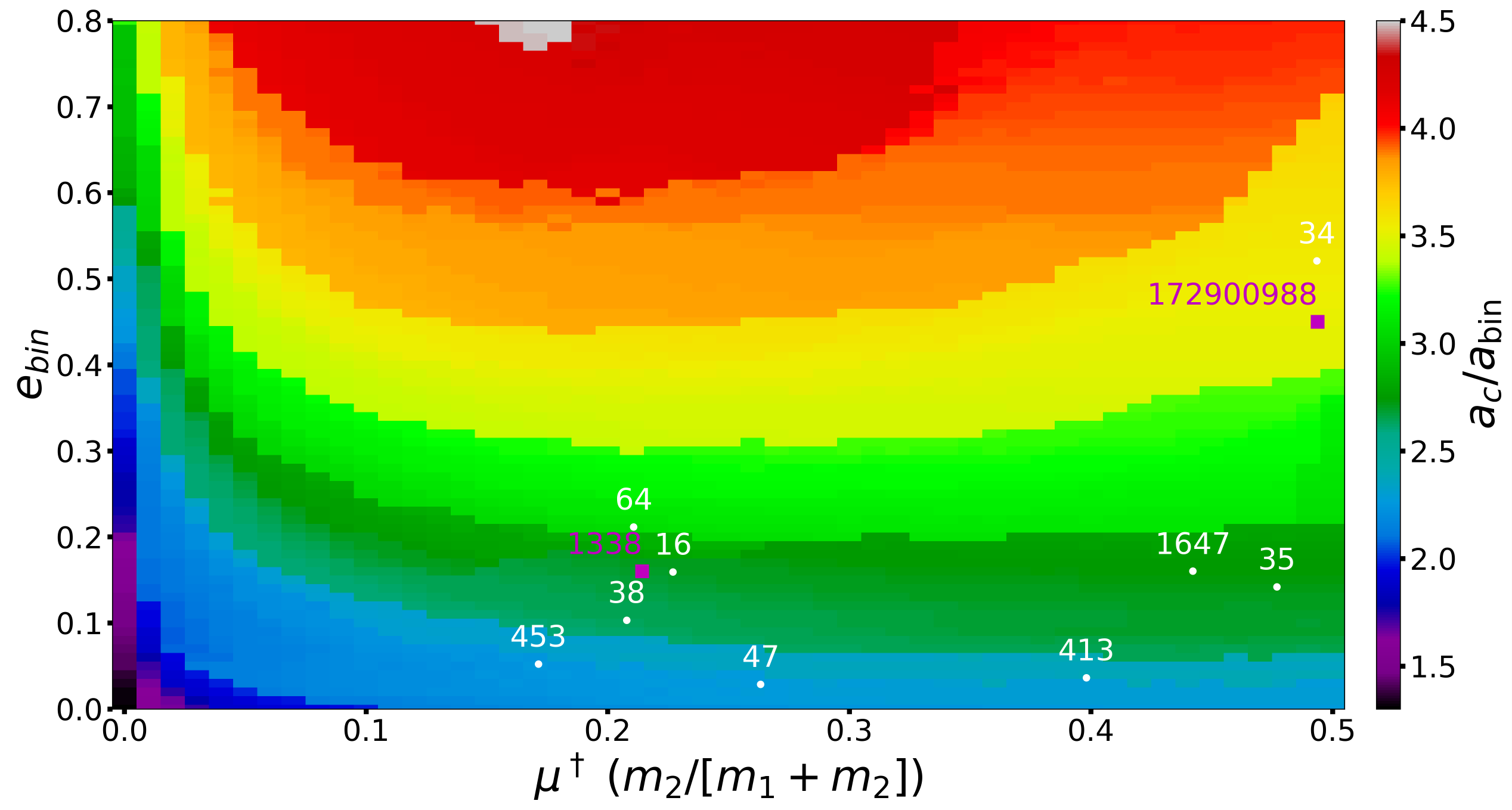}
    \caption{Critical semimajor axis $\alpha_c$ (color-coded) of a P-Type exoplanet as a function of the binary eccentricity $e_{bin}$ and the mass ratio $\mu^\dagger$ ($m_1$ and $m_2$ are the stellar masses). The white dots (and numbers) mark the system parameters for the circumbinary planets discovered using the Kepler Space telescope.  The magenta squares mark the confirmed system (TOI-1338; \cite{Kostov2020}) and planet candidate (TIC 172900988; \cite{Kostov2021}) discovered using TESS \citep{Ricker2015}.  Figure adapted from \citep{Quarles2018b}.}
    \label{fig:CBP_stab}
\end{figure}

For the stability limit in single planet systems, \cite{Quarles2018b} probed to finer ranges in $\mu^\dagger$,  $e_{\rm bin}$, and $\alpha$ compared to \cite{Holman1999}.  This allowed for a revision to the empirical formula for the stability limit through a similar least-squares procedure as \cite{Holman1999}.  The new stability formula is

\begin{align} \label{eqn:Quar_stab_Ptype}
\alpha_c =&\ (1.48 \pm 0.01) + (5.14 \pm 0.10)\mu^\dagger + (3.92 \pm 0.06)e_{\rm bin} + (0.33 \pm 0.19)\mu^\dagger e_{\rm bin} + \nonumber \\ 
&\  (-1.41 \pm 0.06)e_{\rm bin}^2 + (-7.95 \pm 0.15)\left(\mu^\dagger\right)^2 + (-4.89 \pm 0.44)\left(\mu^\dagger e_{\rm bin}\right)^2,
\end{align}

which substantially differs from the coefficients determined by \citeauthor{Holman1999}.  This difference occurs because \cite{Quarles2018b} extended their range in mass ratio down to $\mu^\dagger = 0.001$.  Figure \ref{fig:CBP_stab} illustrates how the critical semimajor axis $\alpha_c$ varies with the binary mass ratio $\mu^\dagger$ and eccentricity $e_{\rm bin}$.  The contours in this map are mostly flat with the binary eccentricity in the middle $(\mu^\dagger = 0.1-0.4)$, where larger deviations occur at the extremes in mass ratio.  This demonstrates that empirical formulas (Eq. \ref{eqn:Hol_stab_Ptype} or \ref{eqn:Quar_stab_Ptype}) will likely be insufficient in the regimes where $\mu^\dagger < 0.1$ or $\mu^\dagger>0.4$.  Figure \ref{fig:CBP_stab} shows that about half of the known circumbinary planets discovered using either the Kepler Space telescope or the Transiting Exoplanet Survey Satellite (TESS; \cite{Ricker2015}) lie within the high mass ratio regime. To avoid this complication, \cite{Quarles2018b} provided a publicly available lookup table\footnote{see the GitHub repo: \href{https://github.com/saturnaxis/CBP_stability}{saturnaxis:CBP\_stability}}, where interpolation routines from \texttt{scipy.interpolate} can be implemented to provide accurate results and potentially go beyond the given grid resolution.  An example is as:

\begin{BoxTypeA}[chap1:box2]{Example 2D interpolation code in \texttt{python}}
\begin{lstlisting}[language=Python]
from scipy.interpolate import CloughTocher2DInterpolator
import numpy as np

mu = 0.05
e_bin = 0.5

Quarles_repo = "https://raw.githubusercontent.com/saturnaxis/CBP_stability/master/"
fn = "a_crit.txt"
X, Y, Z = np.genfromtxt(Quarles_repo + fn,delimiter=',',comments='#',unpack=True)
interp = CloughTocher2DInterpolator(np.array([X,Y]).T, Z)

print("a_c = ",interp(mu,e_bin))
\end{lstlisting}

\end{BoxTypeA}

\cite{Lam2018} used a deep neural network (DNN\footnote{see the GitHub repo: \href{https://github.com/CoolWorlds/orbital-stability}{CoolWorlds:orbital-stability}.}) trained on a million n-body simulations to characterize the stability limit.  The DNN technique provided a machine-learning approach that contrasts to the ${\sim}10^8$ simulations required to produce a general map like Fig. \ref{fig:CBP_stab}.  There are limitations to the DNN approach, where it had a recall accuracy of ${\sim}90\%$ that is likely a trade-off from having a smaller training set.  Using either a lookup table or a DNN is a good first step to estimating the stability limit, where the details (e.g., number of planets, inclined planets, mean motion resonances) are likely to be system specific. More recently, \cite{Georgakarakos24}  revisited the problem of the dynamical stability of hierarchical triple systems with applications to circumbinary planetary orbits. They provided empirical expressions in the form of multidimensional, parameterized fits for the two borders that separate the three dynamical domains, and trained a machine learning model on the data set in order to have an alternative tool of predicting the stability of circumbinary planets.


\section{Conclusion}

This chapter has reviewed the intricate dynamics and stability of planetary systems within binary star environments, an area of study that holds important implications for our understanding of exoplanetary systems. Through a combination of secular methods, N-body simulations, and the application of machine learning techniques, one can analyze the stability of S-type (planets orbiting one star) and P-type (planets orbiting both stars) configurations in binary systems.

Our chapter highlights the critical role of gravitational interactions between stellar components and their orbiting planets, which dictate the long-term stability of these systems. The Laplace-Lagrange framework and disturbing functions have been instrumental in deriving key insights into the secular evolution of planetary orbits \citep[e.g.,][]{Heppenheimer1978,Marchal1990,Moriwaki2004, AndradeInes2017}. By defining critical orbits and stability boundaries, one can identify regions where planets can maintain stable trajectories over extended periods.

The necessity of numerical simulations is underscored by their ability to capture the full complexity of planetary dynamics in binary systems \citep{Rabl1988,Quarles2018a, Quarles2020}. These simulations, augmented by machine learning models, have significantly enhanced our predictive capabilities, allowing for high-accuracy stability assessments \citep{Lam2018} The integration of these advanced methodologies has provided a more comprehensive understanding of the dynamic interactions at play.

As the discovery of exoplanets within binary star systems continues, the frameworks and methods discussed in this chapter will be essential for future research. Understanding the balance of forces that govern planetary stability in these complex environments not only advances our knowledge of celestial mechanics but also informs the characterization of exoplanetary systems.

\bibliographystyle{Harvard}
\bibliography{reference,refStability,hierarchicalMultiples,threebodydynamics}
\end{document}